%% file: main.tex
\documentclass[acmsmall,screen,review=false]{acmart}
\makeatletter
\def\@ACM@checkaffil{% Only warnings
    \if@ACM@instpresent\else
    \ClassWarningNoLine{\@classname}{No institution present for an affiliation}%
    \fi
    \if@ACM@citypresent\else
    \ClassWarningNoLine{\@classname}{No city present for an affiliation}%
    \fi
    \if@ACM@countrypresent\else
        \ClassWarningNoLine{\@classname}{No country present for an affiliation}%
    \fi
}
\makeatother

\pdfoutput=1

\AtBeginDocument{%
  \providecommand\BibTeX{{%
    \normalfont B\kern-0.5em{\scshape i\kern-0.25em b}\kern-0.8em\TeX}}}

\renewcommand\footnotetextcopyrightpermission[1]{}
\usepackage{balance}
\usepackage[utf8]{inputenc}
\usepackage{graphicx}
\usepackage{float}
\usepackage{array}
\usepackage{tabularx}
\usepackage{multirow}
\usepackage{amsmath}

\usepackage{amssymb}
\usepackage[flushleft]{threeparttable}
\usepackage{setspace}
\usepackage{color}
\usepackage{colortbl}
\usepackage{xcolor}
\usepackage{tikzsymbols}
\usepackage{utfsym}
\usepackage{enumitem}
\usepackage{soul}
\usepackage{booktabs}
\usepackage{makecell}
\usepackage{footnote}
\usepackage{algorithm}
\usepackage{algorithmicx}
\usepackage{algpseudocode}
\usepackage{algpseudocode}
\usepackage{amssymb}
\usepackage{hyperref}
\usepackage{caption}
\usepackage{diagbox}
\usepackage{verbatim}
\usepackage[framemethod=TikZ]{mdframed}
\usepackage{tikz}

\newmdenv[
  backgroundcolor=gray!20,
  linewidth=0.5pt,
  roundcorner=6pt,
  tikzsetting={draw=black!70, line width=0.5pt, fill=gray!20},
]{mybox}

\author{Qing~Huang}
\email{qh@whu.edu.cn}
\affiliation{%
  \institution{Jiangxi Normal University, School of Computer Information Engineering}
  \city{Nanchang}
  \state{Jiangxi}
  \country{China}}

\author{Yanbang~Sun}
\authornote{Y. Sun and Q. Huang are co-first authors.}
\affiliation{  
  \institution{Jiangxi Normal University, School of Computer Information Engineering}
  \city{Nanchang}
  \state{Jiangxi}
  \country{China}}
\email{ybsun@jxnu.edu.cn}

\author{Zhenchang Xing}
\affiliation{\institution{CSIRO's Data61}
  \city{Canberra}
  \country{Australia}}
\email{zhenchang.xing@data61.csiro.au}

\author{Yuanlong Cao}
\authornote{Y. Cao is the corresponding author.}
\affiliation{  
  \institution{Jiangxi Normal University, School of Computer Information Engineering}
  \city{Nanchang}
  \state{Jiangxi}
  \country{China}}
\email{ylcao@jxnu.edu.cn}

\author{Jieshan Chen}
\affiliation{\institution{CSIRO's Data61}
  \city{Canberra}
  \country{Australia}}
\email{jieshan.chen@data61.csiro.au}

\author{Xiwei Xu}
\affiliation{\institution{CSIRO's Data61}
  \city{Sydney}
  \country{Australia}}
\email{xiwei.xu@data61.csiro.au}

\author{Huan Jin}
\affiliation{\institution{Jiangxi University of Technology}
  \city{Nanchang}
  \state{Jiangxi}
  \country{China}}
\email{jinhuan@jxut.edu.cn}

\author{Jiaxing Lu}
\affiliation{\institution{Jiangxi Normal University, School of Computer Information Engineering}
  \city{Nanchang}
  \state{Jiangxi}
  \country{China}}
\email{003484@jxnu.edu.cn}

\title{Let's Discover More API Relations: A Large Language Model-based AI Chain for Unsupervised API Relation Inference}

\begin{document}
\begin{sloppypar}

\begin{abstract}
APIs have intricate relations that can be described in text and represented as knowledge graphs to aid software engineering tasks. 
Existing relation extraction methods have limitations, such as limited API text corpus and affected by the characteristics of the input text.
To address these limitations, we propose utilizing large language models (LLMs) (e.g., GPT-3.5) as a neural knowledge base for API relation inference. 
This approach leverages the entire Web used to pre-train LLMs as a knowledge base and is insensitive to the context and complexity of input texts.
To ensure accurate inference, we design our analytic flow as an AI Chain with three AI modules: API FQN Parser, API Knowledge Extractor, and API Relation Decider.
The accuracy of the API FQN parser and API Relation Decider module are 0.81 and 0.83, respectively. 
Using the generative capacity of the LLM and our approach's inference capability, we achieve an average F1 value of 0.76 under the three datasets, significantly higher than the state-of-the-art method's average F1 value of 0.40. 
Compared to CoT-based method, our AI Chain design improves the inference reliability by 67\%, and the AI-crowd-intelligence strategy enhances the robustness of our approach by 26\%.
\end{abstract}


\maketitle

\input{Paper-Section/01.Introduction.tex}

\input{Paper-Section/02.Motivation.tex}
\input{Paper-Section/03.Approach.tex}

\input{Paper-Section/04.Exp_Set.tex}
\input{Paper-Section/05.Exp_Res.tex}

\input{Paper-Section/06.Discussion.tex}
\input{Paper-Section/07.Related_Work.tex}

\input{Paper-Section/08.Conclusion.tex}

\bibliography{sample-base}
\bibliographystyle{unsrt}

\end{sloppypar}
\end{document}

%% file: Paper-Section/01.Introduction.tex
\section{INTRODUCTION}
\begin{figure*}[t]
    \centering
    \includegraphics[width=0.8\textwidth]{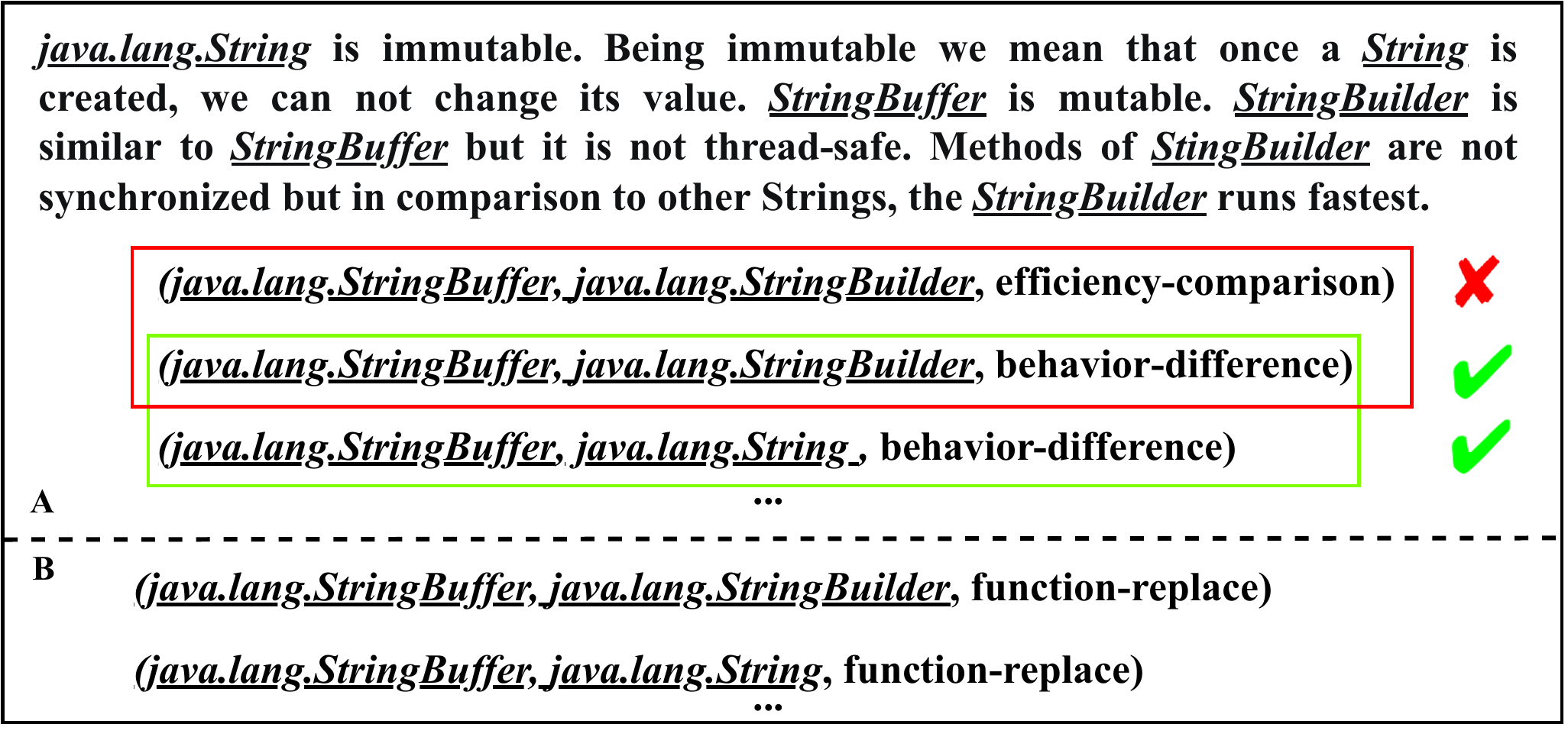}
    \caption{The complexity of API Relations in Text. Part A represents the API relations are explicitly expressed in the text; Part B represents the API relations are not explicitly expressed in the text.}
    \label{fig:Killing example}
 \end{figure*}

Application programming interface (API) related online resource (e.g., Stack Overflow) often contains complex API relations, with one API able to associate with multiple other APIs or multiple relations with another API.
Fig.~\ref{fig:Killing example}-A shows a piece of text from a Stack Overflow post\footnote{\href{https://stackoverflow.com/questions/2971315/string-stringbuffer-and-stringbuilder/24706276}{\textcolor{blue}{https://stackoverflow.com/questions/2971315}}}.
The red box highlights the presence of multiple relations between two APIs, while the green box signifies that a single API has relations with several other APIs.
Specifically, \textit{java.lang.StringBuffer} has efficiency-comparison and behavior-difference relations with \textit{java.lang.StringBuilder}, as well as a behavior-difference relation with \textit{java.lang.String}.
These API relations are summarized by an empirical study~\cite{Huang2022112PK}, and specific definitions can be found in Table~\ref{tab:def_rels}.
They offer great benefits in software development. 
For instance, the behavior-difference relation reminds developers to use the right API for the right task - \textit{StringBuffer} for multi-threading and \textit{StringBuilder} for single-threading. 
The efficiency-comparison relation suggests replacing \textit{StringBuffer} with \textit{StringBuilder} in a single-thread to improve performance.

API relations are described in text, such as API specifications, tutorials, and Q\&A forums. 
Due to the unstructured nature of text, rich API relation knowledge in text is not easily accessible for API recommendation~\cite{Huang2018APIMR}, misuse analysis~\cite{Ren2020APIMisuseDD} and code generation~\cite{Sun2019KnowHowIP}. 
An effective way to improve API relation knowledge accessibility is to extract API relations from text and represent them explicitly as the structural information in a knowledge graph~\cite{Li2018ImprovingAC, Ren2020APIMisuseDD, Liu2020GeneratingCB, Sun2020TaskOrientedAU, Huang2022112PK}.

\begin{table*}[t]
\caption{API Relation Definition}
\label{tab:def_rels}
\label{tab:my-table}
\resizebox{\linewidth}{!}{
% \footnotesize
\begin{tabular}{|l|l|l|}
\hline
Relation               & Definition               & Example  
\\ \hline
Function-Similarity    & Two API entities have similar usage.   & Both java.io.File and java.nio.file can be used for file operations in Java.
\\ \hline
Behavior-Difference    & \begin{tabular}[c]{@{}l@{}}Two API entities behave differently \\ when completing the same task.\end{tabular}  
& \begin{tabular}[c]{@{}l@{}} Java.time.LocalDateTime can represent both date and time, while \\ java.time.LocalTime can only represent time.\end{tabular} 
\\ \hline
Function-Replace       & \begin{tabular}[c]{@{}l@{}} One API entity should be replaced by \\ another API in some specific condition.\end{tabular}   
& \begin{tabular}[c]{@{}l@{}} Database update operations should be executed with \\ Statement.executeUpdate rather than Statement.executeQuery.\end{tabular} 
\\ \hline
Function-Collaboration & \begin{tabular}[c]{@{}l@{}}Two API entities should be used together \\ when  accomplishing a task.\end{tabular}
& \begin{tabular}[c]{@{}l@{}} Using java.io.InputStream and java.io.OutputStream together for file \\ copy operation.\end{tabular}
\\ \hline
Logic-Constraint       & \begin{tabular}[c]{@{}l@{}}One API should be called before or after \\ using another API.\end{tabular} 
& \begin{tabular}[c]{@{}l@{}} You should use java.sql.Connection to establish a connection with the \\ database before using java.sql.Statement.\end{tabular}
\\ \hline
Efficiency-Comparison & \begin{tabular}[c]{@{}l@{}}Two APIs have an efficiency comparison \\ in a  certain condition.\end{tabular}
& \begin{tabular}[c]{@{}l@{}} Java.util.ArrayList is more efficient than java.util.LinkedList for random \\ access of elements.\end{tabular}
\\ \hline
Type-Conversion       &\begin{tabular}[c]{@{}l@{}} Two API entities can be converted to \\ each other.\end{tabular}
& Java.util.stream.Stream and java.util.List can be converted to each other.
\\ \hline
\end{tabular}}
\end{table*}

Existing methods for extracting API relations from unstructured text rely on either hand-crafted sentence patterns~\cite{Li2018ImprovingAC, Huang2022112PK, Huang2018TellTA, Ren2020APIMisuseDD, Sun2020TaskOrientedAU, Liu2020GeneratingCB} or fine-tuning pre-trained language models~\cite{Huang2023APIEA} with labeled sentences. 
However, these approaches have two limitations.

On one hand, existing methods are limited by the scope of the collected API text corpus, which cannot adequately cover diverse API relations. 
Furthermore, existing methods only extract API relations explicitly mentioned from the given text, but cannot make any inference of API relations beyond the provided text corpus.
For example, as shown in Fig.~\ref{fig:Killing example}-B, we are unable to extract the API relation function-replace between \textit{java.lang.StringBuffer} and \textit{java.lang.StringBuilder} (or \textit{java.lang.String}) from the given text.
These API relations can be found in other texts on Stack Overflow\footnote{\href{https://stackoverflow.com/questions/355089/}{\textcolor{blue}{https://stackoverflow.com/questions/355089}}} or other information sources, for example, Java tutorial\footnote{\href{https://www.digitalocean.com/community/tutorials/string-vs-stringbuffer-vs-stringbuilder}{\textcolor{blue}{https://www.digitalocean.com/community/tutorials/string-vs-stringbuffer-vs-stringbuilder}}}.
Unfortunately, sentence patterns are sensitive to description variations across information sources, and data labeling involves significant human effort. 
As a result, expanding existing methods to encompass more information sources is difficult.

On the other hand, the effectiveness of relation extraction methods~\cite{Huang2022112PK,Huang2023APIEA} is affected by the characteristics of the input text, such as syntactic complexity and semantic richness. 
Simple texts, while easier to handle, may not capture diverse API relations.
For example, sentence ``StringBuffer is thread-safe and synchronized whereas StringBuilder is not.'' mentions only behavior-difference relation between \textit{java.lang.StringBuffer} and \textit{java.lang.StringBuilder}, without expressing other API relations such as efficiency-comparison.
Complex texts may express diverse API relations.
However, their high syntactic complexity, involving multiple clauses and pronoun references, poses challenges for relation extraction.
For instance, as shown in Fig.~\ref{fig:Killing example}-A, existing methods~\cite{Huang2022112PK,Huang2023APIEA} can only extract the behavior-difference relation (marked with \usym{2714}) from the text, while overlooking their efficiency-comparison relation (marked with \usym{2718}).

To overcome the two limitations, we propose a novel approach that leverages the large language model (LLM), e.g., GPT 3.5~\cite{GPT-3.5}, as a neural knowledge base for inferring API relations. 
These LLMs are pre-trained on the corpus of the entire Internet (Common Crawl~\cite{Luccioni2021WhatsIT}), which are referred to as foundation models~\cite{bommasani2021opportunities}.
As such foundation LLMs pack the knowledge of the entire Web, we are no longer limited to the scope of API text corpus. 
Our approach does not assume API relations are explicitly mentioned in a piece of text. 
Instead, we consult the LLM for the knowledge (e.g., usage, characteristics, performance) of the two APIs separately and then make the inference of API relations based on the API knowledge. 
As our approach does not rely on the input text for relation inference, it is not sensitive to the characteristics of the input text. 
We rely on the strong generative capability of the LLM to obtain the concise and fluent description of API knowledge for inference. 
This also gives us the capability of inferring more diverse API relations that are not present in the input text.

By leveraging the LLM’s capability for in-context learning, our approach is completely unsupervised, thus removing the needs for developing sentence patterns or data labelling.
The key of our approach is to design an effective analytic flow to interact with the LLM and develop effective prompts to extract API knowledge from the LLM and to make API relation inference.
However, a challenge in this approach is the potential for the LLM to generate incorrect responses to inquires about API relations, especially explicit descriptions of API relations exist but are relatively rare. 
To mitigate this, we do not assume that the LLM can directly answer the inquires about API relations, e.g., “What is the relation between API\_1 and API\_2?” or ``Are API\_1 and API\_2 functionally similar?''.
Instead, we employ Chain of Thought (CoT)~\cite{Wang2022NoMF,Wu2021AICT} to infer API relations, which involves three steps: 
API FQN parsing, API knowledge mining, and API relation decision.

However, the CoT method put all task description within a single prompt, which may lead to an ``epic'' prompt with too many responsibilities and error accumulation.
To address this, we break down the CoT into an AI chain with three modules: API FQN Parser, API Knowledge Extractor, and API Relation Decider, each containing multiple AI units. 
These AI units interact with LLMs step by step to infer API relations. 
We design a prompt with a task instruction and few-shot examples for each AI unit, adapting the LLM to our tasks through in-context learning.
Compared with API FQN Parser and API Knowledge Extractor, API Relation Decider is much more sensitive to prompt engineering. 
To improve the robustness of API relation inference, we design an AI-crowd-intelligence strategy to consult the LLM with prompt variants whose outputs will be combined with the majority vote.

We systematically conduct experiments and analyze the performance of our approach (API Relation Inference, short for APIRI). 
Firstly, we verify the usefulness of two important modules API FQN Parser and API Knowledge Extractor by measuring their accuracy, which are 0.81 and 0.83, respectively. 
Secondly, we compared with the state-of-the-art method (AERJE)~\cite{Huang2023APIEA} in three scenarios, and the results show that the average F1 score of APIRI is 0.76, significantly higher than the average F1 score of 0.40 achieved by AERJE.

Finally, we test the effectiveness of the proposed strategy in our approach. 
The results show that 1) compared with LLM directly answering questions, the analytic flow (CoT-based) improves the response reliability of LLM by 105\%.
2) compared with CoT-based method, AI chain improves the reliability of API relation inference by 67\%.
3) the AI-crowd-intelligence strategy further improved the robustness of our approach by 26\% compared to not adopting this strategy.

In this paper, we make the following contributions:
\begin{itemize}[leftmargin=*]
    \item 
    We propose the use of LLMs as an Internet-scale neural knowledge base and design a LLM-based AI chain and prompts for robust API relation inference.
    \item
    We leverage the LLM’s in-context learning without the need of developing sentence patterns or data labeling.
    \item
    We evaluate the proposed approach, revealing that it outperforms SOTA relation extraction methods in API relation coverage. 
    Our experiments also show that the proposed modular AI chain and AI-crowd-intelligence is effective and robust.  
\end{itemize}

%% file: Paper-Section/02.Motivation.tex
\section{Motivation}
In spite of the hope of obtaining accurate knowledge of API relations from GPT-3.5 after inputting text containing APIs, the LLM more often than not produces incorrect responses~\cite{Ji2022SurveyOH, Bang2023AMM}. 
Two representative cases are presented as follows:

\textbf{Case 1}: When fed with the sentence ``List is an ordered sequence of elements whereas Set is a distinct list of elements which is unordered''\footnote{\href{https://stackoverflow.com/questions/1035008/what-is-the-difference-between-set-and-list}{\textcolor{blue}{https://stackoverflow.com/questions/1035008}}}, GPT-3.5 infers a behavior-difference relation between \textit{List} and \textit{Set}. However, both \textit{List} and \textit{Set} are simple names and may refer to different APIs with fully qualified names (FQNs) such as \textit{java.util.List} or \textit{java.awt.List} for \textit{List} and \textit{java.util.Set} or \textit{org.hibernate.mapping.Set} for \textit{Set}. 
This name ambiguity makes it difficult to accurately determine the relation between the two APIs. 
It's necessary to infer the FQN of the APIs before accurately determining their relations.

\textbf{Case 2}: When fed with the sentence ``If I want a way to get incremental numbers in a multi-threaded environment, should I prefer AtomicInteger.incrementAndGet over AtomicInteger.getAndIncrement''\footnote{\href{https://stackoverflow.com/questions/51402367/duplication-of-number-in-atomicinteger-getandincrement/51442610\#51442610}{\textcolor{blue}{https://stackoverflow.com/questions/51402367}}}, GPT-3.5 infers function-similarity relation between \textit{AtomicInteger.incrementAndGet()} and \textit{AtomicInteger.getAndIncrement()}.
However, a thorough examination reveals that these two APIs have a behavior-difference relation, as the former returns the previous value, while the latter returns the updated value.
This highlights the importance of extracting API knowledge before determining the relation between APIs in a text.

To ensure that the LLM provides accurate knowledge of API relations, it is necessary to first convert the APIs in the text into API FQNs, and then supplement the inference with sufficient API knowledge. 
To achieve this, we decompose the API Relation Inference task into three sub-tasks: parser API FQN, extract API knowledge, and decide API relation. 
Furthermore, these subtasks can be implemented by several functional units, which are categorized as AI units that perform inference based on LLM or Non-AI units that follow explicit predefined rules or logic. 
In summary, the design of these modules/units follows two software engineering principles: separation of concerns and single responsibility, adopting a modular design structure.

%(see Fig.~\ref{fig:framework}-layer 2)

%% file: Paper-Section/03.Approach.tex
\section{APPROACH}
\begin{figure*}[t]
    \centering
    \includegraphics[width=1.0\textwidth]{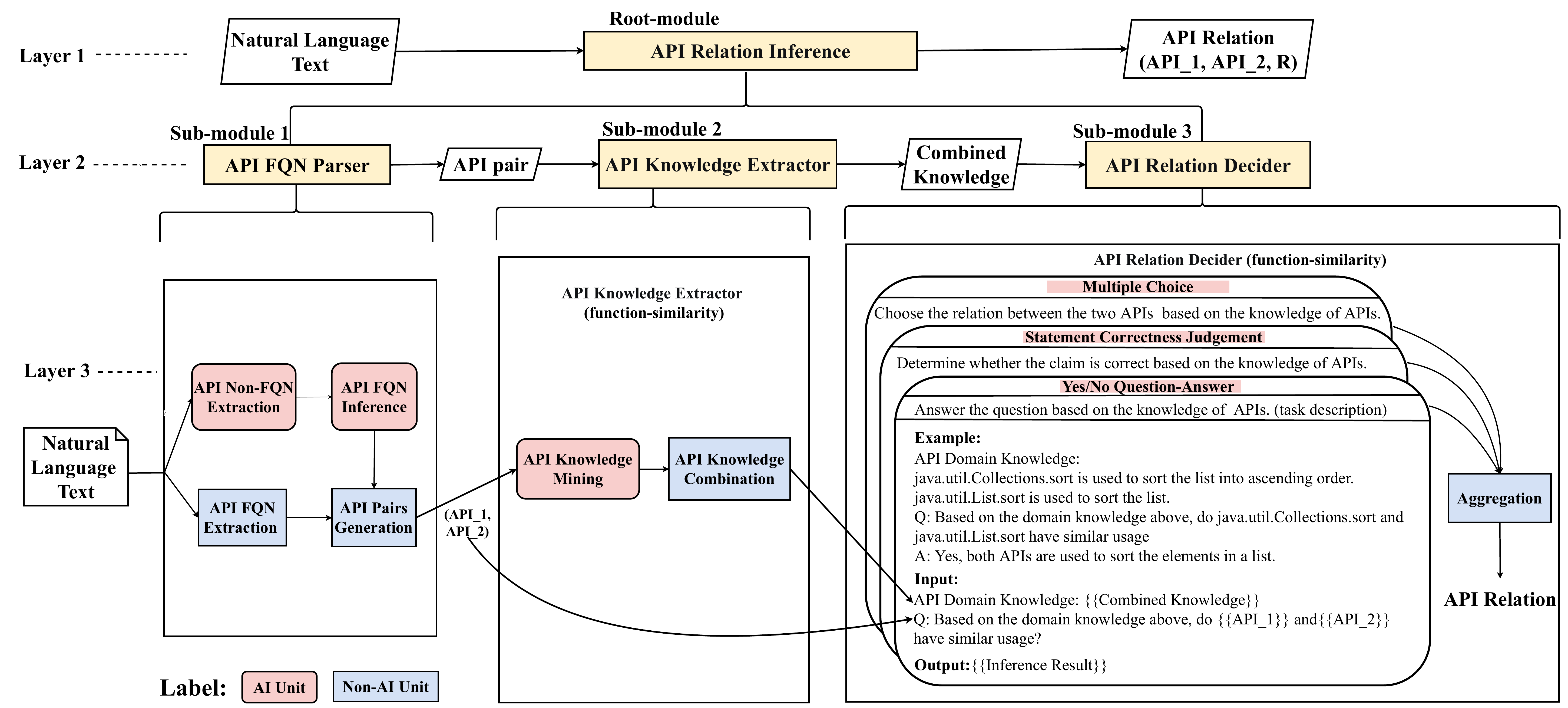}
    \caption{Overall Framework of APIRI (Example for function-similarity)}
    \label{fig:framework}
\end{figure*}

Our APIRI leverages the LLM's in-context learning and its Internet-scale knowledge to infer seven types of API relations: function-similarity, behavior-difference, function-replace, efficiency-comparison, and logic-constraint, function-collaboration, and type-conversion summarized in previous work~\cite{Huang2022112PK, Ren2020APIMisuseDD, Liu2020GeneratingCB}.
It breaks down the task into single-responsibility sub-problems, and designs modules (see Fig.~\ref{fig:framework}-Layer 2) and functional units (see Fig.~\ref{fig:framework}-Layer 3).
These units are then linked in a serial, parallel or split-merge structure to create a multi-round interaction with the LLM to solve problems step by step like a human.
Our current implementation uses GPT-3.5 model (text-davinci-003~\cite{GPT-3.5}) as the underlying LLM.
However, we are not limited to GPT-3.5, but can adopt any LLM with in-context learning capabilities.
Unlike existing machine-learning based methods that require significant effort for data collection, cleaning and labeling as well as model training, evaluation and  optimization, our approach focuses on what problem to solve (task characteristics, data properties, and information flow) by standing on the shoulder of the LLM.
Next, we will introduce the module and unit design of APIRI.

\subsection{API FQN Parser Module}
The API FQN Parser module is designed to identify and extract APIs from text and organize them into pairs of API FQNs.
The parser consists of two AI units (API Non-FQN Extraction and API FQN Inference) and two Non-AI units (API FQN Extraction and API Pairs Generation).
Given a text, it first checks for FQN references and uses the API FQN Extraction unit to extract them through pattern matching proposed in previous API extraction work~\cite{ye2016learning, ye2016software}.
If the text contains Non-FQN references (i.e., simple names), the API Non-FQN Extraction unit is utilized to extract these references and the API FQN Inference unit is used to infer the corresponding FQNs.
The resulting FQNs are then pair-wisely paired by the API Pairs Generation unit, which outputs the API pairs.
For example, API\_1, API\_2, API\_3 can be combined into three API pairs (<API\_1, API\_2>, <API\_1, API\_3>, and <API\_2, API\_3>).
Note that we assume if two APIs are mentioned in a piece of text, the two APIs could be potentially related. 
However, whether a candidate pair of APIs are actually related and what type of relations they have need to be further determined.

\begin{figure}[t]
    \centering
    \includegraphics[width=0.84\textwidth]{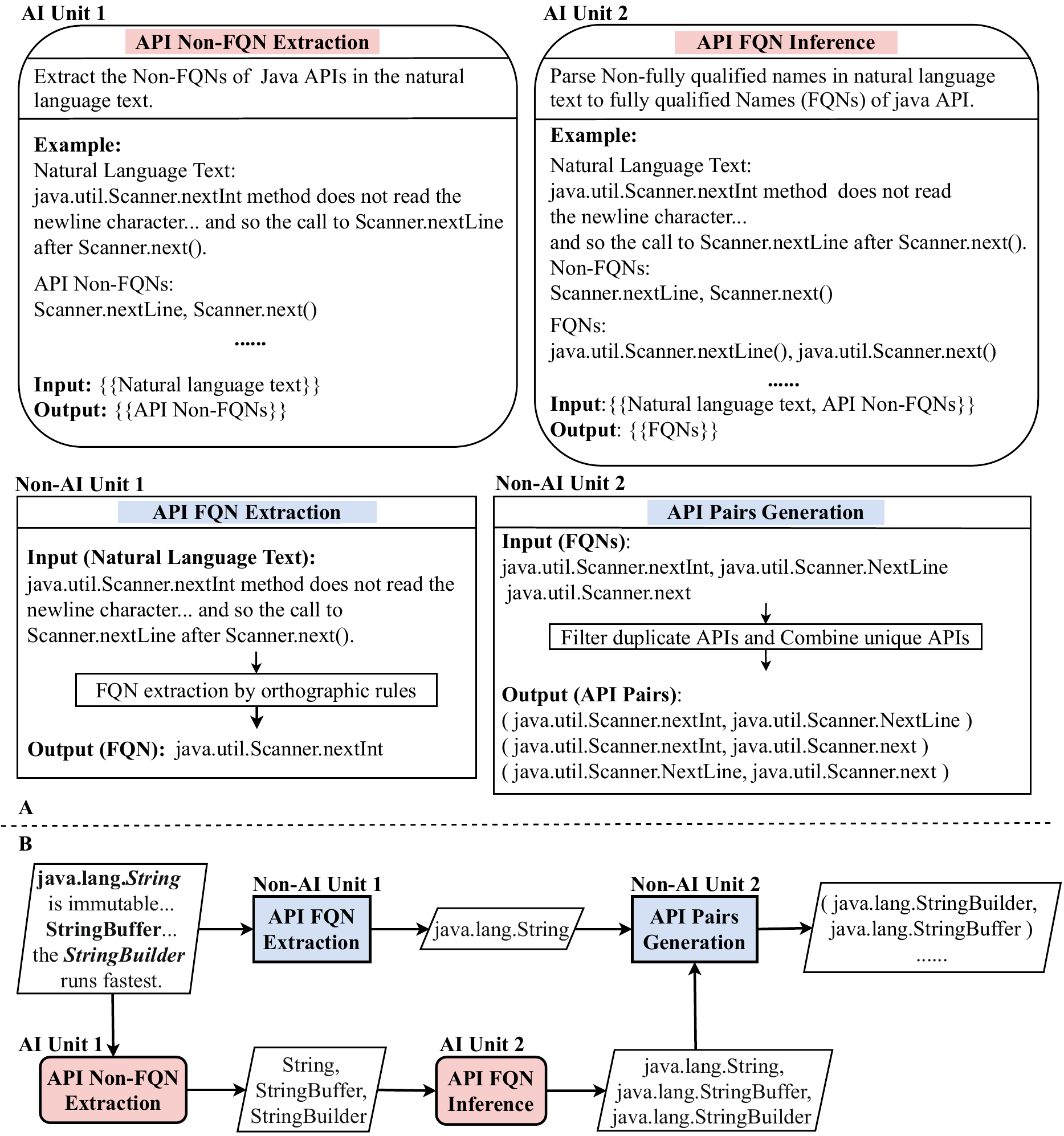}
    \caption{API FQN Parser Module}
    \label{fig:module_1}
\end{figure}

\subsubsection{AI Unit in API FQN Parser}
An empirical study~\cite{Huang2022SEFK} found that descriptions and examples are critical for in-context learning. 
To standardize our prompt design, we develop a generic template that includes a task description and input-output examples. 
The API Non-FQN Extraction unit serves as an example of the template's structure, as shown in Fig.~\ref{fig:module_1}-A.
The template includes a task description (e.g., ``Extract the Non-FQNs...''), followed by five input-output examples (e.g., ``Natural language text: java.util.Scanner.nextInt...'', ``Scanner.nextLine, Scanner.next()''). 
After being provided with a natural language text, API Non-FQN Extraction unit extracts the API Non-FQNs from the text. 
The model adaptability generally increases with more examples~\cite{Huang2022SEFK}, but Min et al.~\cite{min-etal-2022-rethinking} demonstrated that additional shots beyond four examples has limited increase on the accuracy of the multi-choice prompt (i.e., the instruction style of Multiple Choice unit, see Section~\ref{sec: API Relation Decider} for more details).
In this work, we pre-select four examples used for all AI units.
The selection of four examples consider their type and diversity. 
For example, different examples should have different text lengths and expression styles, involve different APIs mentioned in various forms, and APIs involved may or may not have the concerned relation type.

The API FQN Parser module contains two AI units (API Non-FQN Extraction and API FQN Inference). 
These units are designed with accompanying prompts.

\textbf{Prompt Design for API Non-FQN Extraction}. 
This prompt helps extract Non-FQNs (simple names and partially qualified names) of APIs from natural language text. 
As shown in Fig.~\ref{fig:module_1}-A (AI Unit 1), the task description is accompanied by four examples, and a space is provided to enter an API text to get its Non-FQNs. 

\textbf{Prompt Design for API FQN Inference}.
This prompt converts Non-FQNs to FQNs.
As shown in Fig.~\ref{fig:module_1}-A (AI Unit 2), the task description is ``Parse Non-fully qualified name...'', followed by four examples of non-FQNs and their corresponding FQNs.
The Non-FQNs generated by API Non-FQN Extraction unit are appended to the end of the input text to produce the corresponding FQNs.

\subsubsection{Running Example of API FQN Parser}
Fig.~\ref{fig:module_1}-B shows a running example of API FQN Parser.
Given an input text (same as the text in Fig~\ref{fig:Killing example}), it is simultaneously fed into two units: 
API FQN Extraction and API Non-FQN Extraction. 
The former unit extracts FQNs (e.g., \textit{java.lang.String}) from the text, while the latter unit extracts Non-FQNs (e.g., StringBuffer and StringBuilder) present in the text. 
Subsequently, these Non-FQNs are input to the API FQN Inference unit to obtain their corresponding FQNs (e.g., \textit{java.lang.StringBuffer}). 
As a result, we obtain all the API FQNs present in the text. 
Finally, the API Pairs Generation unit combines these API FQNs in pairs to generate several API pairs (e.g., <\textit{java.lang.StringBuffer, java.lang.StringBuilder}>).

\subsection{API Knowledge Extractor Module}
The API Knowledge Extractor module is designed to enhance APIs with specific types of knowledge.
It consists of an AI unit (API Knowledge Mining) and a Non-AI unit (API Knowledge Combination).
Given an API pair, the API Knowledge Mining unit uses the LLM to enrich each API with API knowledge, which is then merged by the API Knowledge Combination unit into a single knowledge block.
Given the need to infer different types of API relations, APIRI has been designed with a separate knowledge extractor for each relation type, all of which operate in parallel.

\subsubsection{AI Unit in API Knowledge Extractor}
We establish an API Knowledge Extractor module for each relation type. 
Each module includes an AI unit named API Knowledge Mining.
As this work considers seven types of API relations, seven prompts are designed, one for each AI unit for a particular relation type, focused on extracting relevant API knowledge for inferring the particular type of relations between API pairs $<$API1, API2$>$.

\begin{figure}[t]
    \centering
    \includegraphics[width=0.92\textwidth]{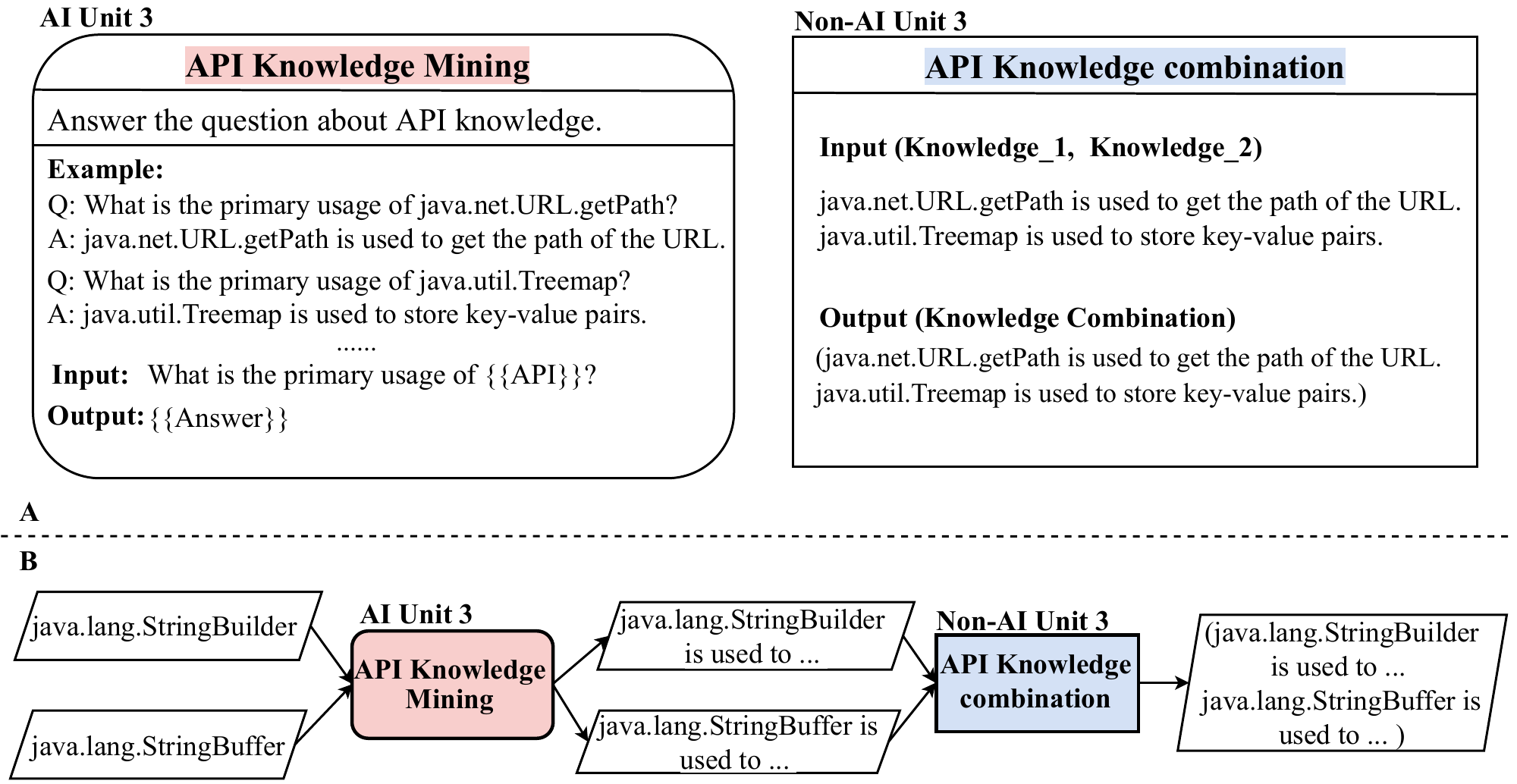}
    \caption{API Knowledge Extractor Module}
    \label{fig:module_2}
\end{figure}

\textbf{Prompt Design for API Knowledge Mining}.
This prompt helps enrich the knowledge of each API in API pair.
Fig.~\ref{fig:module_2}-A (AI Unit 3) shows the API Knowledge Mining unit as an example of extracting API knowledge for inferring function-similarity relations between APIs.
The prompt for the function-similarity relation emphasizes the usage knowledge of each API in the API pair by asking, ``What is the primary usage of \{\{API\}\}?''. 

The other relations are also guided by their relevant knowledge, the details are as follows:
\begin{itemize}[leftmargin=*]
    \item 
    The characteristic knowledge for behavior-difference (``What are the characteristics of \{\{API\}\}?'');
    \item 
    The performance knowledge for efficiency-comparison (``What is the performance of \{\{API\}\}?'');
    \item 
    The condition knowledge for logic-constraint (``What should be done before and after using \{\{API\}\}?''); 
    \item 
    The usage scenario knowledge for function-replace (``When should I use/ not use \{\{API1\}\}?'').
    \item 
    The application scenario knowledge for function-collaboration (``What tasks can \{\{API\}\}  accomplish?'')
    \item 
    The type knowledge for type-conversion (``What data types can \{\{API\}\} be converted to?'')
\end{itemize}

\subsubsection{Running Example of API Knowledge Extractor}
As shown in Fig.~\ref{fig:module_2}-B, the API Knowledge Extractor takes an API pair generated by the API FQN Parser as input (e.g., <\textit{java.lang.StringBuffer, java.lang.StringBuilder}>). 
Each API in the pair is processed in parallel by the API Knowledge Mining unit to obtain its respective usage knowledge (e.g., ``java.lang.StringBuffer is used to...''). 
This parallel execution ensures that in-context learning is not affected by the order of API input, avoiding potential influence from preceding API knowledge. 
Subsequently, the API Knowledge Combination unit merges the individual API knowledge into knowledge block, which are used as input to the API Relation Decider.

\subsection{API Relation Decider Module}
The API Relation Decider module is tasked with determining the relation between APIs.
It consists of three AI units (Yes/No Question-Answer, Statement Correctness Judgement and Multiple Choice), operating in parallel, as well as a Non-AI unit (Result Aggregation).
Given an API pair and its domain knowledge, the three AI units independently assess the relation between the APIs from different perspectives, and their results are then aggregated by the Result Aggregation unit to arrive at a final conclusion.
The goal is to improve the robustness of API relation inference, for which separate deciders have been designed for each type of API relation, operating in parallel.

\subsubsection{AI Unit in API Relation Decider}\label{sec: API Relation Decider}
For each relation type, there is an API Relation Decider module.
Each module has three AI units (API relation decision unit): Yes/No Question-Answer, Statement Correctness Judgement, and Multiple Choice.
Inspired by the different question formats used by Saurav et al. in their work on AI calibrating its own answers~\cite{Kadavath2022LanguageM}, we devise three kinds of instruction styles to improve the reliability of LLM inferring API relations.
As a result, Yes/No Question-Answer, Statement Correctness Judgement, and Multiple Choice employ open-ended question style, restrictive judgment style, and restrictive choice style, respectively, for inferring about API relations.
The prompts for Yes/No Question-Answer and Statement Correctness Judgement units are customized to each relation type, while the prompt for Multiple Choice unit is generic for all relations and displays all relation types as options.

\begin{figure}[t]
    \centering
    \includegraphics[width=0.86\textwidth]{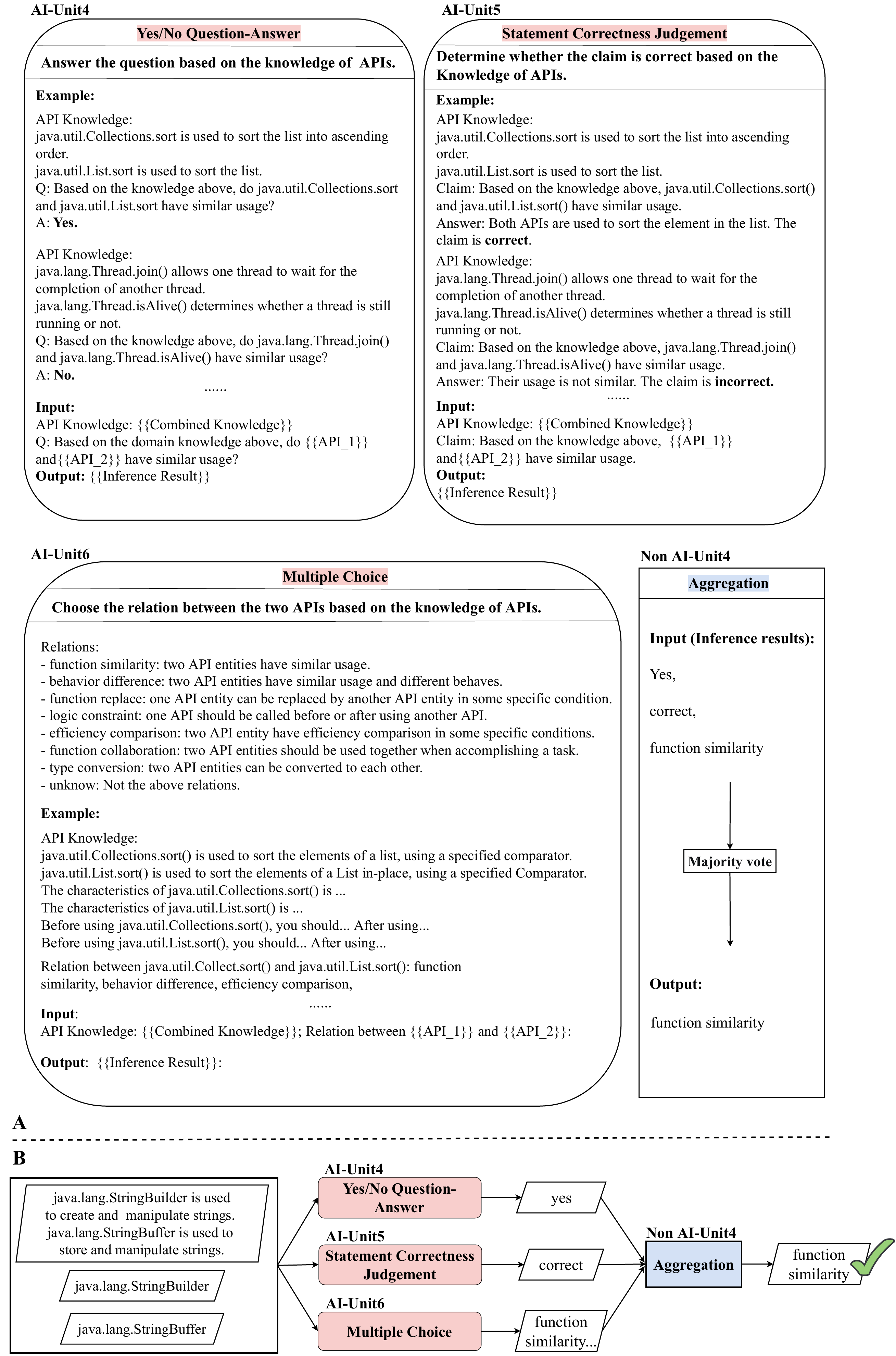}
    \caption{API Relation Decider (function—similarity) Module}
    \label{fig:module_3}
\end{figure}

\clearpage

To enhance the performance of the decider module, we employ a crowd-intelligence strategy using a split-merge structure. 
This strategy enables all three prompts to query the language model concurrently for inferring API relations.
The results are then subjected to a majority vote, providing a more robust outcome.

\textbf{Prompt Design for Yes/No Question-Answer}.
This prompt is used to obtain relation inference results from GPT-3.5 by asking questions.  
As shown in Fig.\ref{fig:module_3}-A (AI Unit 4), the task description is ``answer questions based on the knowledge block of APIs.''
Given the knowledge of two APIs in a pair $<$API1, API2$>$, the prompt asks a question such as ``Based on the knowledge above, do \{\{API1\}\} and \{\{API2\}\} have similar usage?'' and provides four examples with either ``yes''/ ``no'' answers. 
The API pair and related knowledge generated by API FQN Parser and API Knowledge Extractor are then fed into the GPT-3.5 using this prompt format to get the inference result.   

For the other six types of relation inference, the prompt format is similar, but with different questions.
The details are as follows:
\begin{itemize}[leftmargin=*]
    \item 
    The question for behavior-difference relation: ``Do \{\{API1\}\} and \{\{API2\}\} have similar usage and different behaviors?''
    \item 
    The question for efficiency-comparison relation: ``Do \{\{API1\}\} and \{\{API2\}\} have efficiency comparison?''
    \item 
    The question for logic-constraint relation: ``Is there a logical order when using \{\{API1\}\} and \{\{API2\}\}?''
    \item 
    The question for function-repalce relation: ``Can \{\{API1\}\} used in the unavailable of \{\{API2\}\}?'' or ``Can \{\{API2\}\} used in the unavailable of \{\{API1\}\}?''
    \item 
    The question for function-collaboration relation: ``Is there a task that requires \{\{API1\}\} and \{\{API2\}\} to cooperate?''
    \item 
    The question for type-conversion relation: ``Can the data type of \{\{API1\}\} and \{\{API2\}\} be converted to each other?''
\end{itemize}

\textbf{Prompt Design for Statement Correctness Judgement}.
This prompt assesses the validity of claims regarding the relation between two APIs by inputting API pair and related API knowledge into the LLM. 
Fig.~\ref{fig:module_3} (AI Unit 5) shows this prompt, which requires the LLM to determine if the provided claim, such as ``\{\{API1\}\} and \{\{API2\}\} have similar usage,'' is correct or incorrect. 
Again, four examples are provided in the prompt.
The prompts for Yes/No Question-Answer and Statement Correctness Judgement unit differ in one being a question while the other being a declarative claim.
The remaining six types of relation inference have the same prompt format, with only the claimed relation adjusted for each one.

Note that there are three relation inferences in Yes/No Question-Answer and Statement Correctness Judgement units require knowledge obtained from inferring about other relations.
First, in order to judge behavior-difference relation, both usage knowledge and behavioral knowledge are necessary. 
Usage knowledge refers to the API knowledge obtained when mining function-similarity, while behavioral knowledge refers to the API knowledge obtained when mining behavior-difference.
This is because if the functions are not similar, even if the behaviors are different, it is impossible to determine a behavior difference relation between the APIs as differences in behavior can also occur between two unrelated APIs.

Second, to infer logic-constraint relation, both usage knowledge and condition knowledge are necessary. 
Condition knowledge refers to the API knowledge obtained when mining logic-constraint. 
Indeed, this is because the condition knowledge of APIs may mention the requirement of using an API that serves a specific suage, which is presented in the usage knowledge.

Third, usage knowledge and performance knowledge are needed when inferring for efficiency-comparison relation.
This is because if the usage of API are different, the comparison of efficiency is not applicable.
For example, the time complexity of reading a single character from a file using \textit{java.io.FileReader} is O(1), and the time complexity for searching an element in \textit{java.util.HashSet} is O(1).
However, these two APIs have different usage, so there is no efficiency-comparison relation.

\textbf{Prompt Design for Multiple Choice}.
The prompt employed in our approach instructs LLM to select the most accurate relation between two APIs from a set of seven options, along with an ``unknown'' option if no relation can be determined. 
As shown in Fig.~\ref{fig:module_3}-A (AI Unit 6), the prompt has a task description ``choose the relation between...'', seven kinds of API relation definition and four examples of API pairs with their associated knowledge. 
The input to this unit are two APIs and their knowledge block related to the seven types of relations.
The output of this unit is the selected relation based on the input.
Any relation outside the provided options or marked as unknown is considered irrelevant.

\subsubsection{Running Example of API Relation Decider}
Fig.~\ref{fig:module_3}-B illustrates the inference process of the function-similarity relation.
Given two APIs (\textit{java.lang.StringBuilder} and \textit{java.lang.StringBuffer}) and their knowledge blocks (``\textit{java.lang.StringBuilder} is used to...'') as input to three different style of API relation inference units, which output their inference results respectively. 
These results are then fed into an Aggregation unit. 
In this unit, to unify the result format, the outputs of Statement Correctness Judgement and Multiple Choice are mapped. 
That is, if their output is ``correct'' or includes ``function similarity'', it is mapped to ``yes''.
Then, a voting strategy is employed to output the result with the majority of votes (yes), i.e., there is a function-similarity between  \textit{java.lang.StringBuilder} and \textit{java.lang.StringBuffer}.

%% file: Paper-Section/04.Exp_Set.tex
\section{Experiments Setup}
This section begins with three research questions about the performance of our approach and then describes the experimental setup, including baseline, data collection, and evaluation metrics.

\subsection{Research Questions}\label{RQ}
To evaluate the performance of APIRI in API relation inference, we investigate the following research questions:

\begin{itemize}[leftmargin=*]
\item
RQ1: What is the quality of each unit or module in APIRI?
\item
RQ2: How well does APIRI perform in API relation inference?
\item
RQ3: How effective are the task analytic flow and crowd-intelligence strategies employed in APIRI?
\end{itemize}

\subsection{Baselines}\label{baseline_description}
In this work, we evaluate the effectiveness of the overall design of APIRI and its module designs.
We compare APIRI with existing API relation extraction methods to verify the effectiveness of our approach.
There are two main methods for API relation extraction, one is the rule-based method, and the other is the LLM-fine-tuning based method.
The rule-based methods rely on API syntax~\cite{Liu2020GeneratingCB}, special-tag annotated relations~\cite{Li2018ImprovingAC}, or some ad-hoc relation phrases~\cite{Huang2022112PK}.
We get the code of the most recent rule-based method~\cite{Huang2022112PK} on Google\footnote{\href{https://drive.google.com/file/d/1lUhEfS9q8H3IOKU5J-NkU3iRhP6blpL7/view?usp=sharing}{\textcolor{blue}{https://drive.google.com/file/d/1lUhEfS9q8H3IOKU5J-NkU3iRhP6blpL7/view?usp=sharing}}} as a rule-based baseline.
The LLM-fine-tuning based method AERJE~\cite{Huang2023APIEA} includes a BERT-based dynamic prompt generator (i.e., a relation classifier) and a T5-based API entity-relation joint extractor.
We obtain the AERJE's source code and data from Github\footnote{\href{https://github.com/SE-qinghuang/AERJE}{\textcolor{blue}{https://github.com/SE-qinghuang/AERJE}}}, then adopt its classifier training set to fine-tune BERT-based dynamic prompt generator and adopt its final training set to fine-tune T5-based joint extractor.
Finally, we use fine-tuned AERJE to extract API relations.

\begin{figure*}[t]
    \centering
    \includegraphics[width=1.0\textwidth]{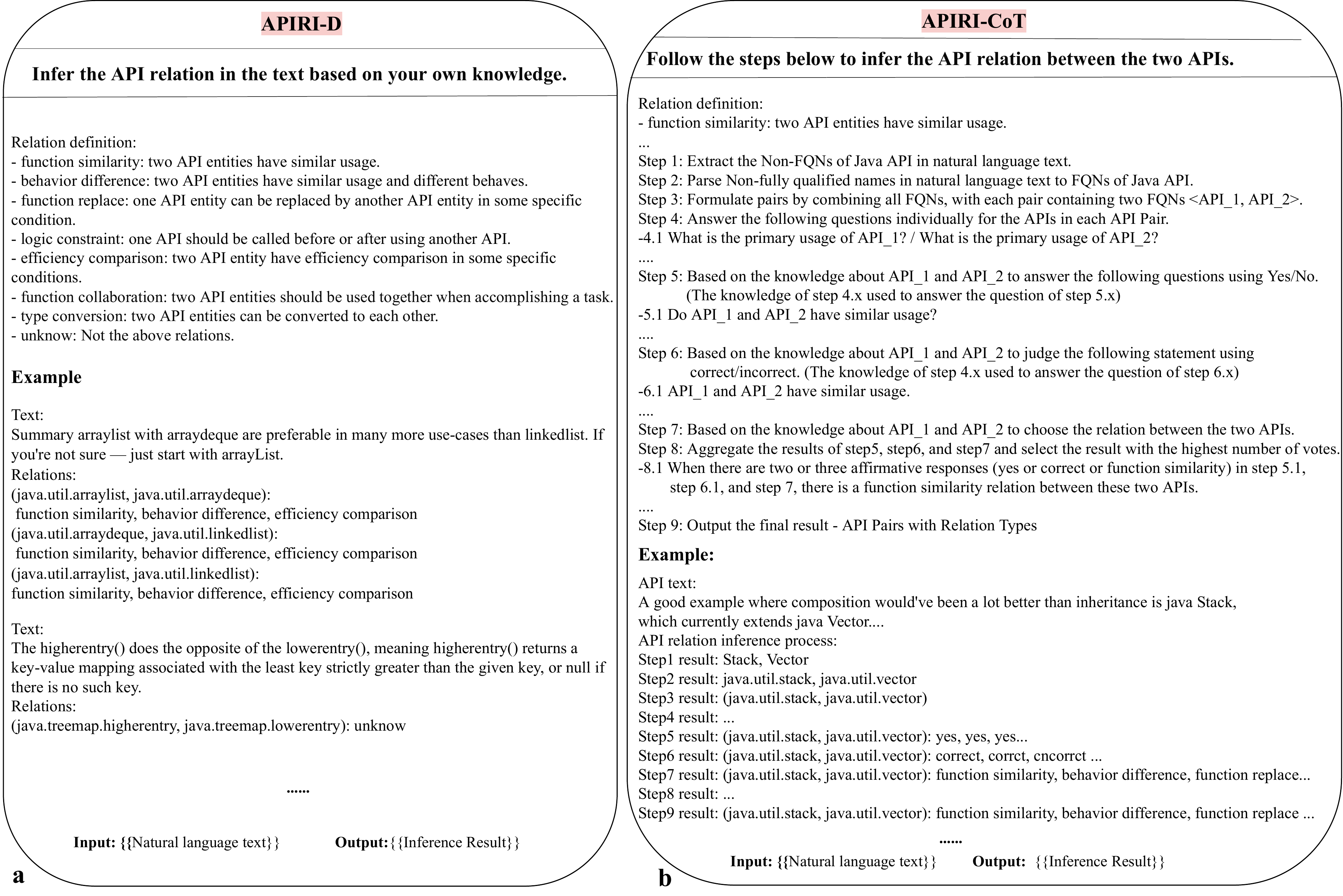}
    \caption{Consult LLM directly (a) and consult LLM based on CoT (b)}
    \label{fig: baseline prompt}
\end{figure*}

We design five variants for the effectiveness and robustness evaluation of APIRI.
One is APIRI-D, which does not contain the idea of chain of thoughts, but directly consults the LLM on what API relations are in the text.
The prompt of APIRI-D can be seen in Fig.~\ref{fig: baseline prompt}-a.
We input a text into APIRI-D, and APIRI-D outputs all the API relations in the text.
APIRI-CoT implements a chain of thought and prompt design, and while it does not use explicit AI modules/units, it describes all steps in an ``epic'' prompt.
As shown in Fig.~\ref{fig: baseline prompt}-b, 
a text is an input into  APIRI-CoT, and the LLM infers the API relation step by step based on the chain of thought we design.
Compared APIRI-D to APIRI-CoT, we can understand the effectiveness of task analytic flow.
By comparing APIRI-CoT with APIRI, we can verify the effectiveness of explicit AI module and unit design in APIRI.
Finally, to evaluate the effect of crowd-intelligence strategy on the robustness of APIRI, we propose three variant methods.
For the API Relation Decider, we devise three variants: ARD-1, ARD-2, ARD-3.
They refer to the API relation decision results using the outputs of the Yes/No Question-Answer, Statement Correctness Judgment, and Multiple Choice units, respectively.

\subsection{Data Preparation}\label{sec:data}
To verify the effectiveness of our approach, we prepare three kinds of datasets, and the specific information is as follows:

\subsubsection{Dataset-1} \label{sec:dataset-1}
To evaluate API relation extraction methods~\cite{Huang2022112PK,Huang2023APIEA} and our approach, we download the test data (they called final test set) of AERJE~\cite{Huang2023APIEA}.
The final test set with a total of 2,686 sentences, 387 of which contain both entities and
relations.
Similar to previous studies~\cite{Li2018ImprovingAC, Wang2019ALA, Liu2020GeneratingCB}, we adopt a sample method~\cite{Singh1996ElementsOS} to ensure that metrics observed in the sample generalize to the population within in a certain confidence interval at a certain confidence level.
For a confidence interval of 5 at a 95\% confidence level, we randomly select 217 texts as test data, we call it \textbf{Dataset-1}.
Due to the AERJE dataset's ground truth being limited to explicit API relations expressed in the text, other API relations beyond the text are not considered. 
As a result, the scope of the API text corpus in AERJE is small. 
Moreover, the dataset consists of individual sentences, which makes the characteristics of API text simple. 
Therefore, we believe that the difficulty of extracting/inferencing API relations from Dataset-1 is low (level-1).
The details of Dataset-1 can be found in the second row of Table~\ref{tab:data_details}.

\subsubsection{Dataset-2}\label{sec:dataset-2}
In order to expand the scope of the API text corpus, we invited 6 PhD students to annotate API relations that are not explicitly expressed in the existing API texts of Dataset-1. 
These students are divided into three groups, with each group annotating the same content independently. 
One author is assigned to resolve conflicts between the students in each group and provide the final answer.
Specifically, we combined the annotated APIs in Dataset-1 in pairs, forming $<$API\_1, API\_2$>$, and then asked the annotators to determine the presence of each API relation in these pairs. 
They are allowed to search for information containing both APIs on Google or retrieve knowledge about each API to decide whether the API relation exists.
The annotation agreement are assessed using the Cohen's Kappa coefficient, which results in a value of 0.82, indicating almost perfect agreement among the annotators.
As a result, \textbf{Dataset-2} is generated, containing a more diverse set of API relations that beyond the explicitly expressed relations in Dataset-1.
However, the text in Dataset-2 still consists of individual sentences, and the characteristics of API text remain simple. 
Therefore, we believe that the difficulty of extracting/inferencing API relations from Dataset-2 is moderate (level-2).
The details of Dataset-2 can be found in the third row of Table~\ref{tab:data_details}.

\subsubsection{Dataset-3}\label{sec:dataset-3}
To expand the scope of the API text corpus and improve the complexity of API text characteristics, we follow the same sampling principles as Dataset-1 and extract 357 API texts from the preprocessed data (Stack Overflow posts) in the AERJE dataset. 
On average, each API text contained 3 sentences. 
These data have not undergone any cleaning process and contain a significant amount of noise. 
Therefore, they are suitable for testing whether APIRI can infer API relations without being influenced by text characteristics.
We then invited 12 PhD students to annotate these API texts.
First, the students annotate the API Non-FQNs and FQNs (Fully Qualified Names) present in the texts, resulting in the ground truth of FQN Parser and its units.
Then, using the same annotation methodology as Dataset-2, these students determine the relations between these APIs. 
This yield the dataset of ground truth for API relations in the text.
We refer to the dataset containing the ground truth for Non-FQN, FQN, and API relations as \textbf{Dataset-3}.
The Kappa coefficient of the annotation process is 0.83 (i.e., almost perfect).
Compared to Dataset-1 and Dataset-2, the extraction or inference of API relations from Dataset-3 presents a high level of difficulty (level-3).
The details of Dataset-3 can be found in the fourth row of Table~\ref{tab:data_details}.

\begin{table}[t]
\caption{Details of each dataset.}
\label{tab:data_details}\
\begin{threeparttable}
\begin{tabular}{|c|c|c|c|c|c|c|c|c|}
\hline
\diagbox[]{DataSet}{Type} & API & f-s & b-d & f-r & e-c & l-c & f-c & t-c \\ \hline
Dataset-1                   & 253 & 42  & 25  & 25  & 14  & 66  & 37  & 8   \\ \hline
Dataset-2                   & 253 & 164 & 151 & 84  & 110 & 115 & 119 & 56  \\ \hline
Dataset-3                   & 569 & 409 & 387 & 113 & 146 & 500 & 531 & 275 \\ \hline
\end{tabular}
\begin{tablenotes}
\small
\item Note: The abbreviations in the table are the first two letters of the API relations.
For example, f-s is short for function-similarity.
\end{tablenotes}
\end{threeparttable}
\end{table}

\subsection{Evaluation Metrics}
Inspired by the measurement of the AI unit method by PCR~\cite{Huang2023PCRChainPC}, we evaluate the AI unit and module's quality (RQ1) using accuracy, which is defined as the ratio of correctly inferred simple names (or FQNs, API relations) to the total number of ground truth for simple names (or FQNs, API relations).
For RQ2 and RQ3, we compare the performance of APIRI and baseline models using Precision, Recall, and F1 metrics, which are used in previous work~\cite{Huang2023APIEA}. 
Precision means what percentage of API relations inferred are correct.
Recall means what percentage of the real API relations are inferred;
F1 score, the harmonic mean of Precision and Recall, is used to measure the overall performance of each method.

%% file: Paper-Section/05.Exp_Res.tex
\section{EXPERIMENTAL RESULTS}
This section delves into three research questions to evaluate and discuss the APIRI’s performance.

\begin{table}[t]
\centering
\caption{The Quality of AI Units and Modules.}
\label{tab:result1}
\begin{tabular}{|c|c|c|c|}
\hline
AI Unit                 & Acc  & Module                                                                          & Acc                   \\ \hline
API Non-FQN Extraction      & 0.83 & \multirow{2}{*}{\begin{tabular}[c]{@{}c@{}}API FQN\\ Parser\end{tabular}}       & \multirow{2}{*}{0.81} \\ \cline{1-2}
API FQN Inference           & 0.89 &                                                                                 &                       \\ \hline
Yes/No Question-Answer & 0.80 & \multirow{3}{*}{\begin{tabular}[c]{@{}c@{}}API Relation\\ Decider\end{tabular}} & \multirow{3}{*}{0.83} \\ \cline{1-2}
Statement Correctness Judgement & 0.78 &                                                                                 &                       \\ \cline{1-2}
Multiple Choice & 0.67 &                                                                                 &                       \\ \hline
\end{tabular}
\end{table}

\subsection{RQ1: What is the Quality of Each Unit or Module in APIRI?}\label{RQ1}

\subsubsection{Motivation}
The CoT approach motivates the decomposition of complex tasks into simpler steps. 
However, the use of a single ``epic'' prompt in CoT-based methods can restrict their effectiveness and result in error accumulation. 
To overcome this limitation, we propose an AI chain with explicit sub-steps, with each step associated with a separate AI unit. 
In this RQ, we evaluate whether each AI unit in our approach can guarantee the accuracy of FQN and API relation inference.

\subsubsection{Methodology}
We apply APIRI to Dataset-3 (i.e., 357 sampled texts) and collect the intermediate results produced by each AI unit.
We then obtain the metrics values (accuracy) by comparing the intermediate results with the ground truth in Dataset-3 (for Non-FQNs, FQNs , and API relations).
Note that in measuring the unit accuracy, we assume independence between units. 
Therefore, when measuring the accuracy of the API FQN inference unit, we use simple names from Dataset-3 (i.e., the ground truth) as input. 
In contrast, for the accuracy of the API FQN Parser module, it reflects the runtime performance of the APIRI.
Thus, the input for API FQN Inference unit should be the output of API Non-FQN Extraction unit. 
We do not evaluate the accuracy of API Knowledge Extractor module as our approach allows this module to be noisy. 
As long as some relevant API knowledge is extracted, the subsequent API Relation Decider module can make inference.
That is, the effect of API knowledge extraction is manifested and evaluated through the performance of API relation decision unit. 

\subsubsection{Result}
The experimental results in Table~\ref{tab:result1} show that most units and modules achieve an accuracy higher than 0.8. 
Specifically, the API FQN Parser module exhibits an accuracy of 0.81, with the API Non-FQN Extraction and API FQN Inference units obtaining accuracy of 0.83 and 0.89, respectively. 
These results demonstrate the AI chain's ability to accurately extract APIs, providing a reliable foundation for API relation inference.
Regarding the API Relation Decider module, although the accuracy of multiple choice is 0.67, the accuracy of the Yes/No Question-Answer unit and the Statement Correctness Judgement unit is 0.80 and 0.78, respectively.
These results show that answering yes/no questions is the most effective to make API relation decisions, while multiple choice is the most challenging mode for the LLM to make correct decisions.
As a result, the API relations obtained through majority voting remain effective.
The overall accuracy of the API Relation Decider module is 0.83, higher than any individual decision units, highlighting the AI chain's capability to accurately infer API relations. 

\vspace{2mm}
\begin{mybox}
\textit{The high accuracy of the AI unit confirms the effectiveness of the proposed prompt design, and the connections of the proposed AI unit can effectively accomplish higher-level tasks.}
\end{mybox}

\subsection{RQ2: How Well Does APIRI Perform in API Relation Inference?}\label{RQ2}
\subsubsection{Motivation}
This RQ aims to verify whether our APIRI, like existing methods~\cite{Huang2022112PK,Huang2023APIEA},
is limited by the scope of API text corpus the characteristics of input API text.

\subsubsection{Methodology}
We apply existing methods~\cite{Huang2022112PK,Huang2023APIEA} and APIRI to three difficulty levels of datasets (Dataset-1, Dataset-2, and Dataset-3). 
The extracted (or inferred) API relations are compared with the ground truth in each dataset, and performance metrics (Precision, Recall, and F1) are calculated for each method.

\subsubsection{Result}
The experimental results are shown in Table~\ref{tab:result2}. 
Overall, APIRI significantly outperforms rule-based methods and AERJE.
The data in Dataset-1 consists of individual sentences, and only the API relations explicitly expressed in the sentences are considered as the ground truth (see Section~\ref{sec:dataset-1}).
In this case, the limitations imposed by the scope of the API text corpus and the characteristics of the API text on the API relation extraction method are weak, making API relation extraction and inference easily achievable.
As demonstrated in the third row of Table~\ref{tab:result2}, AERJE's precision, recall and F1-score are comparable to those of APIRI. 
However, the rule-based method's low recall, due to strict rule matching, results in inferior performance compared to the other two methods.

For Dataset-2, although its data still consists of individual sentences, the ground truth for API relations include not only the API relations explicitly expressed in the sentence, but also the API relations that are not explicitly expressed in the sentence (as stated in Section~\ref{sec:dataset-2}).
In this case, the experimental results of the three methods are shown in the fourth row of Table~\ref{tab:result2}.
Compared to the Rule-based method and AERJE, the F1 score of APIRI is 0.75, significantly higher than the other two methods (0.26, 0.35). 
Furthermore, compared to the experimental results of level-1 (see the third row of Table~\ref{tab:result2}), both the rule-based method and AERJE show a notable decrease (about 50\%) in F1 score, while APIRI's F1 score only decreases by 6.25\%.
This indicates that, unlike existing methods~\cite{Huang2022112PK,Huang2023APIEA}, APIRI can overcome the limitations of the scope of API text corpus.

In Dataset-3, each API text comprises an average of 3 sentences, and the ground truth for API relations includes both the API relations explicitly expressed in the API text and those not explicitly expressed in the API text (as described in Section~\ref{sec:dataset-3}).
In this scenario, the F1 score of Rule-based method and AERJE is only 0.04, while APIRI still keeps a high F1 score (0.72).
Compared to the data of medium difficulty, the F1 score of APIRI only decreased by 4\%. 
However, the F1 scores of existing methods decreased significantly, even exceeding 88\%.
This demonstrates that existing methods are highly sensitive to the complexity of API text characteristics, while APIRI can surpass these limitations.
For example, as shown in Fig.~\ref{fig:Killing example}-A, AERJE can only extract the behavior-difference relation between \textit{StringBuffer} and \textit{StringBuilder}, as well as between \textit{StringBuffer} and \textit{String} from the text, but it overlooks the efficiency-comparison relation between \textit{StringBuffer} and \textit{StringBuilder}.
Furthermore, AERJE is incapable of inferring any other relations (see Fig.~\ref{fig:Killing example}-B) not explicitly expressed in the text.
Instead, APIRI not only accurately infers the efficiency-comparison relation between \textit{StringBuffer} and \textit{StringBuilder} but also has the capability to infer other relations that are not explicitly mentioned in the text, for example, the function-replace relation between \textit{StringBuffer} and \textit{StringBuilder}, as well as between \textit{StringBuffer} and \textit{String}.

\vspace{2mm}
\begin{mybox}
    \textit{Standing on the shoulder of LLM (GPT-3.5), APIRI can accurately infer diverse API relations without being limited by the scope of API text corpus and API text characteristics.}
\end{mybox}

\begin{table}[t]
\centering
\caption{The Results of Two Existing API Relation Extraction Methods and Our APIRI.}
\label{tab:result2}
\begin{threeparttable}
\begin{tabular}{|c|ccl|ccl|ccl|}
\hline
\multirow{2}{*}{\diagbox[]{DataSet}{Method} } & \multicolumn{3}{c|}{Rule-based~\cite{Huang2022112PK}} & \multicolumn{3}{c|}{AERJE~\cite{Huang2023APIEA}} & \multicolumn{3}{c|}{APIRI} \\ \cline{2-10} 
                                               & P         & R        & F1       & P       & R       & F1     & P       & R       & F1     \\ \hline
Dataset-1 (level-1)                            & 0.77      & 0.60    & 0.67     & 0.83    & 0.81    & 0.82   & 0.81    & 0.80    & 0.80   \\ \hline
Dataset-2 (level-2)                            & 0.77      & 0.16     & 0.26    & 0.83    & 0.22    & 0.35   & 0.74    & 0.76    & 0.75   \\ \hline
Dataset-3 (level-3)                            & 0.32      & 0.02     & 0.04     & 0.64    & 0.02    & 0.04   & 0.73    & 0.72    & 0.72   \\ \hline
\end{tabular}
\begin{tablenotes}
\small
\item Note: The higher the level of the dataset, the more complex the data, and the more difficult the extraction and inference of API relations.
\end{tablenotes}
\end{threeparttable}
\end{table}

\subsection{RQ3: How Effective Are the Task Analytic Flow and Crowd-Intelligence Strategies Employed in APIRI?}
\subsubsection{Motivation}
CoT can mitigate the illusion of directly consulting LLMs, but its ``epic'' prompts with too much responsibilities would make CoT-based approaches difficult to control and optimize.
To solve this problem, we designed an AI chain.  
Step by step, the chain interacts with the LLMs to generate the API relation.  
Moreover, to improve the effectiveness of the AI chain, we also design a crowd-intelligence strategy that generates more accurate API relations.
In this RQ, we aim to investigate two aspects of our approach.
Firstly, we would like to explore whether our AI chain design can effectively interact with LLMs.
Secondly, we would like to investigate whether the crowd-intelligence could enhance the robustness of APIRI.

\subsubsection{Methodology}
The five variants of APIRI (as described in Section~\ref{baseline_description}), i.e., APIRI-D, APIRI-CoT, ARD-1, ARD-2, and ARD-3, are applied to Dataset-3.
Compared to Dataset-1 and Dataset-2, Dataset-3 has a higher level of complexity and includes the textual noise present in the real world.
Therefore, testing APIRI with Dataset-3 is more representative.
The inferred API relations are then compared with the ground truth to obtain the metrics (P-R-F1).

\subsubsection{Result}
Table~\ref{tab:result3} presents the experimental results, which show that APIRI outperforms all baseline methods.
Comparing APIRI-D and APIRI-CoT, we observe that the CoT-based approach can improves the response reliability of LLM by 105\%.
This is because if LLM does not see an explicit text describing the relation between the two APIs during its pre-training, it cannot directly infer the API relation.
For instance, the APIRI-D fails to infer the logic-constraint between \textit{java.util.Stream.stream.of} and \textit{java.util.stream.Stream.findfirst}. 
In order to explore whether there is any text containing the relation, we conduct a search on Google for the triplet (\textit{java.util.stream.Stream.of}, \textit{java.util.stream.Stream.findfirst}, logic-constraint).
However, none of the first ten search result pages contained any text describing this API relation.
However, APIRI-CoT can infer the relation by combining these two API's usage knowledge (e.g., ``before using Stream.findfirst() you should create the stream'' and ``Stream.of() is used to create a stream from a given set of elements.'').

\begin{table}[t]
\centering
\caption{Ablation results of APIRI Variants.}
\label{tab:result3}
\begin{threeparttable}
\begin{tabular}{|c|c|c|c|}
\hline
\diagbox[]{Strategy}{Metric} & P                         & R                         & F1                        \\ \hline
APIRI                          & 0.73                      & 0.72                      & 0.72                      \\ \hline
APIRI-D                        & 0.45                      & 0.14                      & 0.21                      \\ \hline
APIRI-CoT                      & 0.65                      & 0.32                      & 0.43                      \\ \hline
ARD-1                          & \multicolumn{1}{l|}{0.67} & \multicolumn{1}{l|}{0.72} & \multicolumn{1}{l|}{0.70} \\ \hline
ARD-2                          & \multicolumn{1}{l|}{0.62} & \multicolumn{1}{l|}{0.77} & \multicolumn{1}{l|}{0.69} \\ \hline
ARD-3                          & \multicolumn{1}{l|}{0.57} & \multicolumn{1}{l|}{0.30} & \multicolumn{1}{l|}{0.39} \\ \hline
\end{tabular}
\end{threeparttable}
\end{table}

As shown in the second and fourth rows of Table~\ref{tab:result3}, the F1 value of APIRI is 0.72, and the F1 value of APIRI-CoT is 0.43, indicating that AI chain improves the reliability of API relation inference by 67\%.
This shows that our AI chain design is superior to CoT's single- prompting approach, which completes all generative steps in a single pass using an ``epic'' prompt with hard-to-control behavior and error accumulation. 
Instead, APIRI breaks down the CoT into an AI chain, with each step corresponding to a separate AI unit that performs separate LLM calls. 
Due to the single responsibility and simplicity of each individual AI unit, the LLM could perform more reliability on individual AI units than on all these steps in a single epic prompt.

Moreover, as shown in the first and last three rows of Table~\ref{tab:result3}, the crowd-intelligence strategy can improve  the robustness of our approach by 26\% on average compared to not using it.
Furthermore, the results of individual API relation decision unit demonstrates that open-ended question style is the most effective of the three prompt styles (open-ended question style, restrictive judgment
style, restrictive choice style) for API relation inference.
This is because answering open-ended questions is consistent with the next token prediction language modeling task~\cite{Arora2022AskMA,ruis2023large}.

\vspace{2mm}
\begin{mybox}
    \textit{AI Chain effectively improves the response reliability of LLM, and the crowd-intelligence strategy further improves the robustness of the approach.
}
\end{mybox}

%% file: Paper-Section/06.Discussion.tex
\section{Threats To Validity}

This section includes three parts: threats to internal validity, threats to external validity and threats to construct validity.

\subsection{Threats to internal validity}
The main internal threat with our approach is error propagation.
Specifically, if an AI unit encounters an error or provides inaccurate results, it can propagate through subsequent units, ultimately leading to incorrect API relations.
Despite the high accuracy of each individual AI unit in our experiments (see Section~\ref{RQ1}), the issue of error propagation persists.
To address the issue of error propagation, we plan to incorporate the scoring and optimization mechanism \cite{fu2023gptscore} into our AI chain in the future.
This mechanism aims to score and optimize the output of each AI unit, minimizing the potential for error propagation and enhancing the effectiveness of our approach.

\subsection{Threats to external validity}
One external threat is our reliance solely on AERJE datasets, which originate from Stack Overflow. 
The labor-intensive data collection and annotation process (though not part of our method) limits our study to the constraint of Stack Overflow text.
Another external threat is this study considers only Java APIs and seven types of API relations.
Furthermore, some overlap exists in the definitions of API relations by Huang et al.~\cite{Huang2022112PK}, for example, logic-constraint is a subset of function-collaboration; behavior difference is a subset of function-similarity. However, to maintain a fair comparison with API relation extraction methods~\cite{Huang2022112PK,Huang2023APIEA}, we have continued to use these definitions without modification.
In the future, we plan to refine the type of API relations to establish clearer boundaries. 
We also intend to expand the scope of our experiments to include more diverse data sources, such as API documentation and programming tutorials, as well as to accommodate other programming languages and other API relation types.
Last but not least, the impact of various prompt content factors (such as instruction style, the number of prompt examples, and the choice of different examples) and the combinations of these factors needs to be further explored.

\subsection{Threats to construct validity}
In this paper, we activate the API knowledge stored in the LLM through in-context learning, without considering the response time of each query. 
In-context learning requires less labeled training data than fine-tuning, thereby saving on manpower and resources. 
However, fine-tuning LLM with more samples can better adapt to specific tasks and generate higher-quality results. 
Fine-tuning can also accelerate response time and reduce latency. 
Thus, if the goal is to obtain superior results, disregarding the cost of resources and manual effort, fine-tuning may be a preferable method to consider.
However, fine-tuned smaller-scale LLM may not pack as broad API knowledge as those LLM pre-trained with Internet-scale text.

Furthermore, in our current implementation, we utilize the GPT-3.5 as the foundation model. 
The improvement in model capabilities (such as upgrading to GPT-4) may affect AI chain and prompt design, but it could have positive effects on directly querying the model, the CoT-based method, and AI chain as well. 
Particularly in the AI chain, prompts with a single responsibility, achieved through task breakdown, might outperform ``epic'' prompts.
Therefore, the performance of AI chain is expected to remain superior to CoT with a single epic-level prompt.
In the future, we plan to conduct experiments using the GPT-4 model.

%% file: Paper-Section/07.Related_Work.tex
\vspace{-3mm}
\section{RELATED WORK}
\vspace{-1mm}
API relations, sourced from API specifications, programming tutorials, and Q\&A forums, offer substantial benefits for software engineering tasks~\cite{Liu2020GeneratingCB, Huang2018TellTA, Ren2020APIMisuseDD, Huang2023APIEA, Li2018ImprovingAC}.
Huang et al.~\cite{Huang2022112PK} have successfully extracted API relations from API documentation to build an API knowledge graph. 
Their research demonstrates that this API knowledge graph significantly improves API search capabilities and enhances developers' understanding of API usage.
Some API relations, such as logic constraint and behavior difference, help prevent API misuse~\cite{Ren2020APIMisuseDD}, while others like efficiency comparison allow developers to write more efficient code~\cite{Liu2020GeneratingCB, Rigby2013DiscoveringEC, Dagenais2012RecoveringTL}.

Early relation extraction techniques relied on human observation of sentence and document structure to create API relation patterns~\cite{Li2018ImprovingAC, Sun2020TaskOrientedAU, Liu2020GeneratingCB,Liu2019GeneratingQC, Ren2020APIMisuseDD, kondoh2019efficient}. 
Recent approaches employ machine learning and human-labeled data to train relation extraction models~\cite{Huang2023APIEA,Gao2019NeuralSF,Zhang2021KnowledgeEnhancedDA}. 
The most advanced method is AERJE~\cite{Huang2023APIEA}, a lightweight joint entity and relation extraction approach that follows the paradigm of pre-train, prompt-tuning, and predict.
In contrast, our approach utilizes unsupervised in-context learning on a pre-trained language model~\cite{bommasani2021opportunities}. 
Furthermore, existing methods extract API relations from limited text sources and only extract relations explicitly stated in the text. 
In contrast, our approach leverages the extensive knowledge packed in the language model to infer API relations that are not explicitly stated in the input text.

In-context learning is a novel paradigm that enables the adaptation of foundation models to new tasks through zero- or few-shot prompts, without gradient updates~\cite{raffel2020exploring, Radford2019LanguageMA, Brown2020LanguageMA}. 
This paradigm has been successfully applied in various software engineering tasks, including program repair~\cite{joshi2022repair, xia2022practical, Kang2022LargeLM}, testing~\cite{Chen2022CodeTCG}, code generation~\cite{Mastropaolo2023OnTR}, and GUI automation~\cite{Liu2022FillIT}.
% These works use a coarse-grained, direct-inquiry style prompt design, similar to our APIRI-D baseline. 
Researchers investigate effective prompt formats and designs~\cite{gao2020making, Mishra2021ReframingIP, Wang2022SelfConsistencyIC, Huang2022SEFK, rubin2021learning, bach2022promptsource, schick2021few, tam2021improving, scao2021many} which inspire our prompt designs.
Especially, chain of thoughts (CoT)~\cite{Wu2021AICT, wei2022chain} has been proposed to address the LLM's limitations for reasoning complex tasks.
However, existing CoT works provides only a simple instruction like ``let’s do something step by step''.
In contrast, our approach distinguishes itself by performing explicit task workflow analysis and modular design, creating an AI chain that interacts with the language model in explicit steps. 
While the idea of AI chains has been explored in writing assistants and question-answering tasks~\cite{coenen2021wordcraft,lee2022coauthor, chung2022talebrush, Wu2021AICT, Schick2022PEERAC,Gordon2017Re3R}, our AI chain involves much more complex task structure and data flow for a domain-specific knowledge inference task.

Supervised prompt-tuning has demonstrated strong few-shot learning capability~\cite{Wang2022NoMF,Luo2022PRCBERTPL,XingSaner2023, han2022ptr, lester2021power, li2021prefix, liu2021p, schick2020exploiting, Huang2022PrompttunedCL} by aligning the learning objectives of downstream tasks and pre-training through prompts.
Huang et al.~\cite{Huang2022PrompttunedCL} uses supervised prompt-tuning for inferring API FQNs in partial code.
Our API FQN Inference unit infers API FQNs in text.
Some works extract factual knowledge from the LLM using a fill-in-blank template~\cite{heinzerling-inui-2021-language, Heinzerling2021LanguageMA,Wang2020LanguageMA,devlin2018bert,cohen-etal-2023-crawling}.
Our AI units extract API knowledge by open-ended question answering.
Finally, as existing supervised prompt-tuning methods are task-specific, they cannot handle a complex task like API relation inference which involve different types of information/knowledge and mix of AI and non-AI units.

%% file: Paper-Section/08.Conclusion.tex
\section{CONCLUSIONS AND FUTURE WORK}
This paper presents a novel approach for inferring intricate API relations using a large language model (LLM) as a neural knowledge base.  
Our approach offers several advantages, including unsupervised in-context learning for API relation inference and the utilization of the entire Web as a knowledge base.  
To enhance the reliability and robustness of the LLM's responses, we design an analytic flow based on software engineering principles and employ effective prompt engineering practices, supported by an AI-crowd-intelligence strategy.
Our approach outperforms state-of-the-art relation extraction methods and achieves high accuracy in inferring API relations of various types.  
In the future, we plan to extend our methodology to other software engineering tasks such as testing, program repair, and bug analysis.

%% file: main.bbl
\begin{thebibliography}{10}

\bibitem{Huang2022112PK}
Qing Huang, Yuan Zhiqiang, Zhenchang Xing, Zhengkang Zuo, Changjing Wang, and
  Xin Xia.
\newblock 1+1$>$2: Programming know-what and know-how knowledge fusion,
  semantic enrichment and coherent application.
\newblock {\em IEEE Transactions on Services Computing}, 2022.

\bibitem{Huang2018APIMR}
Qiao Huang, Xin Xia, Zhenchang Xing, D.~Lo, and Xinyu Wang.
\newblock Api method recommendation without worrying about the task-api
  knowledge gap.
\newblock {\em 2018 33rd IEEE/ACM International Conference on Automated
  Software Engineering (ASE)}, pages 293--304, 2018.

\bibitem{Ren2020APIMisuseDD}
Xiaoxue Ren, Xinyuan Ye, Zhenchang Xing, Xin Xia, Xiwei Xu, Liming Zhu, and
  Jianling Sun.
\newblock Api-misuse detection driven by fine-grained api-constraint knowledge
  graph.
\newblock {\em 2020 35th IEEE/ACM International Conference on Automated
  Software Engineering (ASE)}, pages 461--472, 2020.

\bibitem{Sun2019KnowHowIP}
Jiamou Sun, Zhenchang Xing, Rui Chu, Heilai Bai, Jinshui Wang, and Xin Peng.
\newblock Know-how in programming tasks: From textual tutorials to
  task-oriented knowledge graph.
\newblock {\em 2019 IEEE International Conference on Software Maintenance and
  Evolution (ICSME)}, pages 257--268, 2019.

\bibitem{Li2018ImprovingAC}
Hongwei Li, Sirui Li, Jiamou Sun, Zhenchang Xing, Xin Peng, Mingwei Liu, and
  Xuejiao Zhao.
\newblock Improving api caveats accessibility by mining api caveats knowledge
  graph.
\newblock {\em 2018 IEEE International Conference on Software Maintenance and
  Evolution (ICSME)}, pages 183--193, 2018.

\bibitem{Liu2020GeneratingCB}
Yang Liu, Mingwei Liu, Xin Peng, Christoph Treude, Zhenchang Xing, and Xiaoxin
  Zhang.
\newblock Generating confcept based api element comparison using a knowledge
  graph.
\newblock {\em 2020 35th IEEE/ACM International Conference on Automated
  Software Engineering (ASE)}, pages 834--845, 2020.

\bibitem{Sun2020TaskOrientedAU}
Jiamou Sun, Zhenchang Xing, Xin Peng, Xiwei Xu, and Liming Zhu.
\newblock Task-oriented api usage examples prompting powered by programming
  task knowledge graph.
\newblock {\em 2021 IEEE International Conference on Software Maintenance and
  Evolution (ICSME)}, pages 448--459, 2020.

\bibitem{Huang2018TellTA}
Yi~Huang, Chunyang Chen, Zhenchang Xing, Tian Lin, and Yang Liu.
\newblock Tell them apart: Distilling technology differences from crowd-scale
  comparison discussions.
\newblock {\em 2018 33rd IEEE/ACM International Conference on Automated
  Software Engineering (ASE)}, pages 214--224, 2018.

\bibitem{Huang2023APIEA}
Qing Huang, Yanbang Sun, Zhenchang Xing, Min Yu, Xiwei Xu, and Qinghua Lu.
\newblock Api entity and relation joint extraction from text via dynamic
  prompt-tuned language model.
\newblock {\em ACM Trans. Softw. Eng. Methodol.}, 2023.

\bibitem{GPT-3.5}
OpenAI.
\newblock Openai gpt-3.5 model.
\newblock \url{https://platform.openai.com/docs/models/gpt-3-5}.
\newblock Accessed: 2023.2.

\bibitem{Luccioni2021WhatsIT}
Alexandra~Sasha Luccioni and Joseph~D. Viviano.
\newblock What’s in the box? an analysis of undesirable content in the common
  crawl corpus.
\newblock In {\em Annual Meeting of the Association for Computational
  Linguistics}, 2021.

\bibitem{bommasani2021opportunities}
Rishi Bommasani, Drew~A Hudson, Ehsan Adeli, Russ Altman, Simran Arora, Sydney
  von Arx, Michael~S Bernstein, Jeannette Bohg, Antoine Bosselut, Emma
  Brunskill, et~al.
\newblock On the opportunities and risks of foundation models.
\newblock {\em arXiv preprint arXiv:2108.07258}, 2021.

\bibitem{Wang2022NoMF}
Chaozheng Wang, Yuanhang Yang, Cuiyun Gao, Yun Peng, Hongyu Zhang, and
  Michael~R. Lyu.
\newblock No more fine-tuning? an experimental evaluation of prompt tuning in
  code intelligence.
\newblock {\em Proceedings of the 30th ACM Joint European Software Engineering
  Conference and Symposium on the Foundations of Software Engineering}, 2022.

\bibitem{Wu2021AICT}
Tongshuang~Sherry Wu, Michael Terry, and Carrie~J. Cai.
\newblock Ai chains: Transparent and controllable human-ai interaction by
  chaining large language model prompts.
\newblock {\em Proceedings of the 2022 CHI Conference on Human Factors in
  Computing Systems}, 2021.

\bibitem{Ji2022SurveyOH}
Ziwei Ji, Nayeon Lee, Rita Frieske, Tiezheng Yu, Dan Su, Yan Xu, Etsuko Ishii,
  Yejin Bang, Wenliang Dai, Andrea Madotto, and Pascale Fung.
\newblock Survey of hallucination in natural language generation.
\newblock {\em ACM Computing Surveys}, 55:1 -- 38, 2022.

\bibitem{Bang2023AMM}
Yejin Bang, Samuel Cahyawijaya, Nayeon Lee, Wenliang Dai, Dan Su, Bryan Wilie,
  Holy Lovenia, Ziwei Ji, Tiezheng Yu, Willy Chung, Quyet~V. Do, Yan Xu, and
  Pascale Fung.
\newblock A multitask, multilingual, multimodal evaluation of chatgpt on
  reasoning, hallucination, and interactivity.
\newblock {\em ArXiv}, 2023.

\bibitem{ye2016learning}
Deheng Ye, Zhenchang Xing, Chee~Yong Foo, Jing Li, and Nachiket Kapre.
\newblock Learning to extract api mentions from informal natural language
  discussions.
\newblock In {\em 2016 IEEE International Conference on Software Maintenance
  and Evolution (ICSME)}, pages 389--399. IEEE, 2016.

\bibitem{ye2016software}
Deheng Ye, Zhenchang Xing, Chee~Yong Foo, Zi~Qun Ang, Jing Li, and Nachiket
  Kapre.
\newblock Software-specific named entity recognition in software engineering
  social content.
\newblock In {\em 2016 IEEE 23rd international conference on software analysis,
  evolution, and reengineering (SANER)}, volume~1, pages 90--101. IEEE, 2016.

\bibitem{Huang2022SEFK}
Qing Huang, Dianshu Liao, Zhenchang Xing, Zhiqiang Yuan, Qinghua Lu, Xiwei Xu,
  and Jiaxing Lu.
\newblock Se factual knowledge in frozen giant code model: A study on fqn and
  its retrieval.
\newblock {\em ArXiv}, abs/2212.08221, 2022.

\bibitem{min-etal-2022-rethinking}
Sewon Min, Xinxi Lyu, Ari Holtzman, Mikel Artetxe, Mike Lewis, Hannaneh
  Hajishirzi, and Luke Zettlemoyer.
\newblock Rethinking the role of demonstrations: What makes in-context learning
  work?
\newblock In {\em Proceedings of the 2022 Conference on Empirical Methods in
  Natural Language Processing}, pages 11048--11064. Association for
  Computational Linguistics, December 2022.

\bibitem{Kadavath2022LanguageM}
Saurav Kadavath, Tom Conerly, Amanda Askell, T.~J. Henighan, Dawn Drain, Ethan
  Perez, Nicholas Schiefer, Zachary Dodds, Nova DasSarma, Eli Tran-Johnson,
  Scott Johnston, Sheer El-Showk, Andy Jones, Nelson Elhage, Tristan Hume, Anna
  Chen, Yuntao Bai, Sam Bowman, Stanislav Fort, Deep Ganguli, Danny Hernandez,
  Josh Jacobson, John Kernion, Shauna Kravec, Liane Lovitt, Kamal Ndousse,
  Catherine Olsson, Sam Ringer, Dario Amodei, Tom~B. Brown, Jack Clark,
  Nicholas Joseph, Benjamin Mann, Sam McCandlish, Christopher Olah, and Jared
  Kaplan.
\newblock Language models (mostly) know what they know.
\newblock {\em ArXiv}, 2022.

\bibitem{Wang2019ALA}
Chong Wang, Xin Peng, Mingwei Liu, Zhenchang Xing, Xue Bai, Bing Xie, and Tuo
  Wang.
\newblock A learning-based approach for automatic construction of domain
  glossary from source code and documentation.
\newblock {\em Proceedings of the 2019 27th ACM Joint Meeting on European
  Software Engineering Conference and Symposium on the Foundations of Software
  Engineering}, 2019.

\bibitem{Singh1996ElementsOS}
Ravindra~Pal Singh and Naurang~Singh Mangat.
\newblock Elements of survey sampling.
\newblock 1996.

\bibitem{Huang2023PCRChainPC}
Qing Huang, Jiahui Zhu, Zhilong Li, Zhenchang Xing, Changjing Wang, and Xiwei
  Xu.
\newblock Pcr-chain: Partial code reuse assisted by hierarchical chaining of
  prompts on frozen copilot.
\newblock In {\em 2023 IEEE/ACM 45th International Conference on Software
  Engineering: Companion Proceedings (ICSE-Companion)}, pages 1--5, 2023.

\bibitem{Arora2022AskMA}
Simran Arora, Avanika Narayan, Mayee~F Chen, Laurel Orr, Neel Guha, Kush
  Bhatia, Ines Chami, and Christopher Re.
\newblock Ask me anything: A simple strategy for prompting language models.
\newblock 2023.

\bibitem{ruis2023large}
Laura~Eline Ruis, Akbir Khan, Stella Biderman, Sara Hooker, Tim
  Rockt{\"a}schel, and Edward Grefenstette.
\newblock Large language models are not zero-shot communicators, 2023.

\bibitem{fu2023gptscore}
Jinlan Fu, See-Kiong Ng, Zhengbao Jiang, and Pengfei Liu.
\newblock Gptscore: Evaluate as you desire.
\newblock {\em arXiv preprint arXiv:2302.04166}, 2023.

\bibitem{Rigby2013DiscoveringEC}
Peter~C. Rigby and Martin~P. Robillard.
\newblock Discovering essential code elements in informal documentation.
\newblock {\em 2013 35th International Conference on Software Engineering
  (ICSE)}, pages 832--841, 2013.

\bibitem{Dagenais2012RecoveringTL}
Barth{\'e}l{\'e}my Dagenais and Martin~P. Robillard.
\newblock Recovering traceability links between an api and its learning
  resources.
\newblock {\em 2012 34th International Conference on Software Engineering
  (ICSE)}, pages 47--57, 2012.

\bibitem{Liu2019GeneratingQC}
Mingwei Liu, Xin Peng, Andrian Marcus, Zhenchang Xing, Wenkai Xie, Shuangshuang
  Xing, and Yang Liu.
\newblock Generating query-specific class api summaries.
\newblock {\em Proceedings of the 2019 27th ACM Joint Meeting on European
  Software Engineering Conference and Symposium on the Foundations of Software
  Engineering}, 2019.

\bibitem{kondoh2019efficient}
Yushi Kondoh, Masashi Nishimoto, Keiji Nishiyama, Hideyuki Kawabata, and Tetsuo
  Hironaka.
\newblock Efficient searching for essential api member sets based on inclusion
  relation extraction.
\newblock {\em International Journal of Networked and Distributed Computing},
  7(4):149--157, 2019.

\bibitem{Gao2019NeuralSF}
Tianyu Gao, Xu~Han, Ruobing Xie, Zhiyuan Liu, Fen Lin, Leyu Lin, and Maosong
  Sun.
\newblock Neural snowball for few-shot relation learning.
\newblock In {\em AAAI Conference on Artificial Intelligence}, 2019.

\bibitem{Zhang2021KnowledgeEnhancedDA}
Jiawen Zhang, Jiaqi Zhu, Yi~Yang, Wandong Shi, Congcong Zhang, and Hongan Wang.
\newblock Knowledge-enhanced domain adaptation in few-shot relation
  classification.
\newblock {\em Proceedings of the 27th ACM SIGKDD Conference on Knowledge
  Discovery \& Data Mining}, 2021.

\bibitem{raffel2020exploring}
Colin Raffel, Noam Shazeer, Adam Roberts, Katherine Lee, Sharan Narang, Michael
  Matena, Yanqi Zhou, Wei Li, Peter~J Liu, et~al.
\newblock Exploring the limits of transfer learning with a unified text-to-text
  transformer.
\newblock {\em J. Mach. Learn. Res.}, 21(140):1--67, 2020.

\bibitem{Radford2019LanguageMA}
Alec Radford, Jeff Wu, Rewon Child, David Luan, Dario Amodei, and Ilya
  Sutskever.
\newblock Language models are unsupervised multitask learners.
\newblock 2019.

\bibitem{Brown2020LanguageMA}
Tom~B. Brown, Benjamin Mann, Nick Ryder, Melanie Subbiah, Jared Kaplan,
  Prafulla Dhariwal, Arvind Neelakantan, Pranav Shyam, Girish Sastry, Amanda
  Askell, Sandhini Agarwal, Ariel Herbert-Voss, Gretchen Krueger, Tom Henighan,
  Rewon Child, Aditya Ramesh, Daniel~M. Ziegler, Jeffrey Wu, Clemens Winter,
  Christopher Hesse, Mark Chen, Eric Sigler, Mateusz Litwin, Scott Gray,
  Benjamin Chess, Jack Clark, Christopher Berner, Sam McCandlish, Alec Radford,
  Ilya Sutskever, and Dario Amodei.
\newblock Language models are few-shot learners.
\newblock 2020.

\bibitem{joshi2022repair}
Harshit Joshi, Jos{\'e} Cambronero, Sumit Gulwani, Vu~Le, Ivan Radicek, and
  Gust Verbruggen.
\newblock Repair is nearly generation: Multilingual program repair with llms.
\newblock {\em arXiv preprint arXiv:2208.11640}, 2022.

\bibitem{xia2022practical}
Chunqiu~Steven Xia, Yuxiang Wei, and Lingming Zhang.
\newblock Practical program repair in the era of large pre-trained language
  models.
\newblock {\em arXiv preprint arXiv:2210.14179}, 2022.

\bibitem{Kang2022LargeLM}
Sungmin Kang, Juyeon Yoon, and Shin Yoo.
\newblock Large language models are few-shot testers: Exploring llm-based
  general bug reproduction.
\newblock {\em ICSE}, 2023.

\bibitem{Chen2022CodeTCG}
Bei Chen, Fengji Zhang, A.~Nguyen, Daoguang Zan, Zeqi Lin, Jian-Guang Lou, and
  Weizhu Chen.
\newblock Codet: Code generation with generated tests.
\newblock {\em ICLR}, 2023.

\bibitem{Mastropaolo2023OnTR}
Antonio Mastropaolo, Luca Pascarella, Emanuela Guglielmi, Matteo Ciniselli,
  Simone Scalabrino, Rocco Oliveto, and Gabriele Bavota.
\newblock On the robustness of code generation techniques: An empirical study
  on github copilot.
\newblock 2023.

\bibitem{Liu2022FillIT}
Zhe Liu, Chunyang Chen, Junjie Wang, Xing Che, Yuekai Huang, Jun Hu, and Qing
  Wang.
\newblock Fill in the blank: Context-aware automated text input generation for
  mobile gui testing.
\newblock {\em ArXiv}, abs/2212.04732, 2022.

\bibitem{gao2020making}
Tianyu Gao, Adam Fisch, and Danqi Chen.
\newblock Making pre-trained language models better few-shot learners.
\newblock In {\em Proceedings of the 59th Annual Meeting of the Association for
  Computational Linguistics and the 11th International Joint Conference on
  Natural Language Processing (Volume 1: Long Papers)}, 2021.

\bibitem{Mishra2021ReframingIP}
Swaroop Mishra, Daniel Khashabi, Chitta Baral, Yejin Choi, and Hannaneh
  Hajishirzi.
\newblock Reframing instructional prompts to gptk’s language.
\newblock In {\em Findings}, 2021.

\bibitem{Wang2022SelfConsistencyIC}
Xuezhi Wang, Jason Wei, Dale Schuurmans, Quoc~V Le, Ed~H. Chi, Sharan Narang,
  Aakanksha Chowdhery, and Denny Zhou.
\newblock Self-consistency improves chain of thought reasoning in language
  models.
\newblock In {\em The Eleventh International Conference on Learning
  Representations}, 2023.

\bibitem{rubin2021learning}
Ohad Rubin, Jonathan Herzig, and Jonathan Berant.
\newblock Learning to retrieve prompts for in-context learning.
\newblock {\em Proceedings of the 2022 Conference of the North American Chapter
  of the Association for Computational Linguistics}, 2022.

\bibitem{bach2022promptsource}
Stephen~H Bach, Victor Sanh, Zheng-Xin Yong, Albert Webson, Colin Raffel,
  Nihal~V Nayak, Abheesht Sharma, Taewoon Kim, M~Saiful Bari, Thibault Fevry,
  et~al.
\newblock Promptsource: An integrated development environment and repository
  for natural language prompts.
\newblock {\em Proceedings of the 60th Annual Meeting of the Association for
  Computational Linguistics System Demonstrations}, 2022.

\bibitem{schick2021few}
Timo Schick and Hinrich Sch{\"u}tze.
\newblock Few-shot text generation with natural language instructions.
\newblock In {\em Proceedings of the 2021 Conference on Empirical Methods in
  Natural Language Processing}, pages 390--402, 2021.

\bibitem{tam2021improving}
Derek Tam, Rakesh R.~Menon, Mohit Bansal, Shashank Srivastava, and Colin
  Raffel.
\newblock Improving and simplifying pattern exploiting training.
\newblock In {\em Proceedings of the 2021 Conference on Empirical Methods in
  Natural Language Processing}, 2021.

\bibitem{scao2021many}
Teven~Le Scao and Alexander~M Rush.
\newblock How many data points is a prompt worth?
\newblock {\em Proceedings of the 2021 Conference of the North American Chapter
  of the Association for Computational Linguistics: Human Language
  Technologies}, 2021.

\bibitem{wei2022chain}
Jason Wei, Xuezhi Wang, Dale Schuurmans, Maarten Bosma, Fei Xia, Ed~Chi, Quoc~V
  Le, Denny Zhou, et~al.
\newblock Chain-of-thought prompting elicits reasoning in large language
  models.
\newblock {\em Advances in Neural Information Processing Systems},
  35:24824--24837, 2022.

\bibitem{coenen2021wordcraft}
Ann Yuan, Andy Coenen, Emily Reif, and Daphne Ippolito.
\newblock Wordcraft: Story writing with large language models.
\newblock In {\em 27th International Conference on Intelligent User
  Interfaces}, 2022.

\bibitem{lee2022coauthor}
Mina Lee, Percy Liang, and Qian Yang.
\newblock Coauthor: Designing a human-ai collaborative writing dataset for
  exploring language model capabilities.
\newblock In {\em Proceedings of the 2022 CHI Conference on Human Factors in
  Computing Systems}, pages 1--19, 2022.

\bibitem{chung2022talebrush}
John Joon~Young Chung, Wooseok Kim, Kang~Min Yoo, Hwaran Lee, Eytan Adar, and
  Minsuk Chang.
\newblock Talebrush: sketching stories with generative pretrained language
  models.
\newblock In {\em Proceedings of the 2022 CHI Conference on Human Factors in
  Computing Systems}, pages 1--19, 2022.

\bibitem{Schick2022PEERAC}
Timo Schick and Hinrich Sch{\"u}tze.
\newblock Exploiting cloze-questions for few-shot text classification and
  natural language inference.
\newblock In {\em Proceedings of the 16th Conference of the European Chapter of
  the Association for Computational Linguistics: Main Volume}, 2021.

\bibitem{Gordon2017Re3R}
Daniel Gordon, Ali Farhadi, and Dieter Fox.
\newblock Re3 : Real-time recurrent regression networks for object tracking.
\newblock {\em ArXiv}, abs/1705.06368, 2017.

\bibitem{Luo2022PRCBERTPL}
Xianchang Luo, Yinxing Xue, Zhenchang Xing, and Jiamou Sun.
\newblock Prcbert: Prompt learning for requirement classification using
  bert-based pretrained language models.
\newblock {\em Proceedings of the 37th IEEE/ACM International Conference on
  Automated Software Engineering}, 2022.

\bibitem{XingSaner2023}
Xiangwei Li, Xiaoning Ren, Yinxing Xue, Zhenchang Xing, and Jiamou Sun.
\newblock Prediction of vulnerability characteristics based on vulnerability
  description and prompt learning.
\newblock In {\em 2023 IEEE International Conference on Software Analysis,
  Evolution and Reengineering (SANER)}, pages 604--615, 2023.

\bibitem{han2022ptr}
Xu~Han, Weilin Zhao, Ning Ding, Zhiyuan Liu, and Maosong Sun.
\newblock Ptr: Prompt tuning with rules for text classification.
\newblock {\em AI Open}, 3:182--192, 2022.

\bibitem{lester2021power}
Brian Lester, Rami Al-Rfou, and Noah Constant.
\newblock The power of scale for parameter-efficient prompt tuning.
\newblock In {\em Proceedings of the 2021 Conference on Empirical Methods in
  Natural Language Processing}, 2021.

\bibitem{li2021prefix}
Xiang~Lisa Li and Percy Liang.
\newblock Prefix-tuning: Optimizing continuous prompts for generation.
\newblock In {\em Proceedings of the 59th Annual Meeting of the Association for
  Computational Linguistics and the 11th International Joint Conference on
  Natural Language Processing (Volume 1: Long Papers)}, 2021.

\bibitem{liu2021p}
Xiao Liu, Kaixuan Ji, Yicheng Fu, Weng Tam, Zhengxiao Du, Zhilin Yang, and Jie
  Tang.
\newblock {P}-tuning: Prompt tuning can be comparable to fine-tuning across
  scales and tasks.
\newblock In {\em Proceedings of the 60th Annual Meeting of the Association for
  Computational Linguistics (Volume 2: Short Papers)}, 2022.

\bibitem{schick2020exploiting}
Timo Schick and Hinrich Sch{\"u}tze.
\newblock Exploiting cloze questions for few shot text classification and
  natural language inference.
\newblock {\em arXiv preprint arXiv:2001.07676}, 2020.

\bibitem{Huang2022PrompttunedCL}
Qing Huang, Zhiqiang Yuan, Zhenchang Xing, Xiwei Xu, Liming Zhu, and Qinghua
  Lu.
\newblock Prompt-tuned code language model as a neural knowledge base for type
  inference in statically-typed partial code.
\newblock {\em International Conference on Automated Software
  Engineering(ASE)}, 2022.

\bibitem{heinzerling-inui-2021-language}
Benjamin Heinzerling and Kentaro Inui.
\newblock Language models as knowledge bases: On entity representations,
  storage capacity, and paraphrased queries.
\newblock In {\em Proceedings of the 16th Conference of the European Chapter of
  the Association for Computational Linguistics: Main Volume}, 2021.

\bibitem{Heinzerling2021LanguageMA}
Benjamin Heinzerling and Kentaro Inui.
\newblock Language models as knowledge bases: On entity representations,
  storage capacity, and paraphrased queries.
\newblock {\em arXiv preprint arXiv:2008.09036}, 2020.

\bibitem{Wang2020LanguageMA}
Chenguang Wang, Xiao Liu, and Dawn~Xiaodong Song.
\newblock Language models are open knowledge graphs.
\newblock {\em ArXiv}, abs/2010.11967, 2020.

\bibitem{devlin2018bert}
Jacob Devlin, Ming-Wei Chang, Kenton Lee, and Kristina Toutanova.
\newblock {BERT}: Pre-training of deep bidirectional transformers for language
  understanding.
\newblock In {\em Proceedings of the 2019 Conference of the North {A}merican
  Chapter of the Association for Computational Linguistics: Human Language
  Technologies, Volume 1 (Long and Short Papers)}, 2019.

\bibitem{cohen-etal-2023-crawling}
Roi Cohen, Mor Geva, Jonathan Berant, and Amir Globerson.
\newblock Crawling the internal knowledge-base of language models.
\newblock In {\em Findings of the Association for Computational Linguistics:
  EACL 2023}, pages 1856--1869. Association for Computational Linguistics,
  2023.

\end{thebibliography}
